\documentclass[longauth]{aa} 

\usepackage{graphicx}
\usepackage{txfonts}
\usepackage{multirow}
\usepackage{chemformula}
\usepackage{hyperref}
\hypersetup{colorlinks=true, linkcolor=red,citecolor=blue,urlcolor=magenta} 

\usepackage{orcidlink}

\usepackage{comment}
\usepackage{booktabs}
\usepackage{float}
\usepackage{tikz}
\usepackage{placeins}  
\usepackage{multicol}

\newcommand{\met}{CH$_3$OH\xspace}
\newcommand{\metc}{CH$_3$CN\xspace}

\newcommand{\et}{C$_2$H$_5$OH\xspace}

\newcommand{\dme}{CH$_3$OCH$_3$\xspace}
\newcommand{\mf}{CH$_3$OCHO\xspace}

\newcommand{\ad}{CH$_3$CHO\xspace}
\newcommand{\fmm}{NH$_2$CHO\xspace}
\newcommand{\2}{$_2$}

\newcommand{\kms}{\,km\,s$^{-1}$\xspace}

\begin{document} 

   \title{PRODIGE -- envelope to disk with NOEMA} 
   \subtitle{VII. (Complex) organic molecules in the NGC1333\,IRAS\,4B1 outflow: \\ A new laboratory for shock chemistry}

  \author{L.~A.~Busch
          \inst{\ref{MPE}}\fnmsep\thanks{\url{lbusch@mpe.mpg.de}}\orcidlink{0009-0009-0269-0491}
          \and
          J.~E.~Pineda \inst{\ref{MPE}}\orcidlink{0000-0002-3972-1978}
          \and
          P.~Caselli \inst{\ref{MPE}}\orcidlink{0000-0003-1481-7911}
          \and
          D.~M.~Segura-Cox \inst{\ref{ROC},\ref{MPE}}\orcidlink{0000-0003-3172-6763}
          \and
          S.~Narayanan\inst{\ref{MPE}}\orcidlink{0000-0002-0244-6650}
          \and
          C.~Gieser \inst{\ref{MPIA}}\orcidlink{0000-0002-8120-1765}
         \and 
          M.~J.~Maureira \inst{\ref{MPE}}\orcidlink{0000-0002-7026-8163}
          \and 
          T.-H. Hsieh \inst{\ref{TARA},\ref{AST},\ref{MPE}}\orcidlink{0000-0002-5507-5697}
         \and 
         Y.~Lin\inst{\ref{MPE}}\orcidlink{0000-0001-9299-5479}
          \and
          M.~T.~Valdivia-Mena \inst{\ref{ESO},\ref{MPE}}\orcidlink{0000-0002-0347-3837}
         \and 
         L.~Bouscasse \inst{\ref{IRAM}}
         \and
          Th.~Henning \inst{\ref{MPIA}}
          \and 
          D.~Semenov \inst{\ref{UH},\ref{MPIA}}\orcidlink{0000-0002-3913-7114}
          \and 
          A.~Fuente\inst{\ref{CAB}}\orcidlink{0000-0001-6317-6343}
          \and
          Y.-R.~Chou\inst{\ref{MPE}}\orcidlink{0009-0004-9608-6132}
          \and
          L.\,Mason\inst{\ref{MPE}}\orcidlink{0009-0005-9178-6751}
          \and 
          P.~C.~Cort\'es\inst{\ref{JAO},\ref{NRAO}}\orcidlink{0000-0002-3583-780X}
          \and
          L.~W.~Looney\inst{\ref{DAI}}\orcidlink{0000-0002-4540-6587}
          \and
          I.~W.~Stephens\inst{\ref{DEEP}}\orcidlink{0000-0003-3017-4418}
          \and
          M.~Tafalla\inst{\ref{IGN}}
          \and
          A.~Dutrey\inst{\ref{LAB}}\orcidlink{0000-0003-2341-5922}
          \and
          W.~Kwon\inst{\ref{DSK},\ref{SNUK}}\orcidlink{0000-0003-4022-4132}
          \and
          P.~Saha\inst{\ref{AST}}\orcidlink{0000-0002-0028-1354}
          }

   \institute{Max-Planck-Institut f\"ur extraterrestrische Physik, Gie\ss enbachstra\ss e 1, 85748 Garching bei M\"unchen, Germany \label{MPE}
        \and
        Department of Physics and Astronomy, University of Rochester, Rochester, NY 14627, USA \label{ROC}
        \and
        Max-Planck-Institut f\"{u}r Astronomie, K\"{o}nigstuhl 17, D-69117 Heidelberg, Germany \label{MPIA}
       \and
       Taiwan Astronomical Research Alliance (TARA), Taiwan \label{TARA}
       \and
       Institute of Astronomy and Astrophysics, Academia Sinica, P.O. Box 23-141, Taipei 106, Taiwan \label{AST}
       \and
       European Southern Observatory, Karl-Schwarzschild-Stra\ss e 2, 85748 Garching, Germany \label{ESO}
       \and 
       Institute de Radioastronomie Millim\'{e}trique (IRAM), 300 rue de la Piscine, F-38406, Saint-Martin d'H\`{e}res, France \label{IRAM}
       \and Zentrum f\"{u}r Astronomie der Universit\"{a}t Heidelberg, Institut f\"{u}r Theoretische Astrophysik, Albert-Ueberle-Str. 2, 69120 Heidelberg, Germany \label{UH}
       \and
       Centro de Astrobiolog\'{i}a (CAB), CSIC-INTA, Ctra. de Ajalvir Km. 4, 28850, Torrej\'{o}n de Ardoz, Madrid, Spain \label{CAB}
       \and
       Joint ALMA Observatory, Alonso de C\'ordova 3107, Vitacura, Santiago, Chile \label{JAO}
       \and
       National Radio Astronomy Observatory, 520 Edgemont Road, Charlottesville, VA 22903, USA \label{NRAO}
       \and
       Department of Astronomy, University of Illinois, 1002 W Green St., Urbana, IL 61801, USA \label{DAI}
       \and
       Department of Earth, Environment and Physics, Worcester State University, Worcester, MA 01602, USA \label{DEEP}
       \and
       Observatorio Astron\'{o}mico Nacional (IGN), Alfonso XII 3, E-28014, Madrid, Spain \label{IGN}
       \and
       Laboratoire d'Astrophysique de Bordeaux, Universit\'{e} de Bordeaux, CNRS, B18N, All\'{e}e Geoffroy Saint-Hilaire, F-33615 Pessac, France\label{LAB}
       \and
       Department of Earth Science Education, Seoul National University, 1 Gwanak-ro, Gwanak-gu, Seoul 08826, Republic of Korea \label{DSK}
       \and
       SNU Astronomy Research Center, Seoul National University, 1 Gwanak-ro, Gwanak-gu, Seoul 08826, Republic of Korea \label{SNUK}
        }

   \date{Received ; accepted }

  \abstract  
  {Shock chemistry is an excellent tool to shed light on the formation and destruction mechanisms of complex organic molecules (COMs). The L1157-mm outflow is the only low-mass protostellar outflow that has extensively been studied in this regard.}
   {We aim to map COM emission and derive the molecular composition of the protostellar outflow driven by the Class\,0 protostar NGC\,1333 IRAS\,4B1 to introduce it as a new laboratory to study the impact of shocks on COM chemistry.
    } 
   {We used the data taken as part of the PRODIGE (PROtostars \& DIsks: Global Evolution) large program to compute integrated intensity maps of outflow emission to identify spatial differences between species. The emission spectra were then analysed towards two positions, one in each outflow lobe, by deriving synthetic spectra and population diagrams assuming conditions of local thermodynamic equilibrium (LTE).}
   {In addition to typical outflow tracers such as SiO and CO, outflow emission is seen from H\2CO, HNCO, and HC\3N, as well as from the COMs \met, \metc, and \ad, and even from deuterated species such as DCN, D\2CO, and CH\2DOH. Maps of integrated intensity ratios between \met and DCN, D\2CO, and \ad reveal gradients with distance from the protostar. For DCN and D\2CO, this may reflect their prestellar abundance profile, provided the outflow is young enough, while an explanation is  still required for \ad.
   Intensity ratio maps of HC\3N and \metc with respect to \met peak in the southern lobe where temperatures are highest. This may point to enhanced HC\3N and \metc formation at this location, potentially in the warmer gas phase.
   Rotational temperatures are found in the range $\sim$50\,--100\,K, which is, on average, warmer than for the L1157-B1 shock spot ($\lesssim$\,30\,K). Abundances with respect to \met are higher by factors of a few than for L1157-B1.}
   {For the first time, we securely detected COMs \metc, \ad, and CH\2DOH in the IRAS\,4B1 outflow, serendipitously with limited sensitivity and bandwidth. Targeted observations will enable the discovery of new COMs and a more detailed analysis of their emission. Morphological differences between molecules in the IRAS\,4B1 outflow lobes and their relative abundances provide first proof that this outflow is a promising new laboratory for shock chemistry, which will offer crucial information on COM formation and destruction as well as outflow structure and kinematics. }

    \authorrunning{L.A.~Busch et al.}
    \titlerunning{}
   \maketitle
 
\section{Introduction}\label{s:intro}

One of the key questions in astrochemistry is to understand the growth of molecular complexity during the process of star formation, including the formation and destruction processes of complex organic molecules \citep[COMs; $\geq$\,6 atoms and carbon-bearing;][]{Herbst09}.
In general, COMs can either form in the gas phase or in the solid phase on icy dust grain surfaces, from which they can desorb through numerous thermal and non-thermal processes \citep[e.g. see review on COM chemistry by][]{Ceccarelli2023}.
To shed light on COM formation and destruction, observing sites of shock chemistry have become increasingly popular.
Shock waves passing through quiescent medium greatly impact the local chemical composition \citep[e.g.][]{Draine1995}. 
They compress and temporally heat the medium, for example, enabling endothermic reactions to take place in the gas phase and sublimating ice mantles, thereby, releasing molecules to the gas phase. Stronger shocks may even sputter or completely destroy grains \citep[e.g.][]{Lenzuni1995,Jimenez-serra2008,Gusdorf2008}. These processes happen on a rather short timescale \citep[100s to 1000s of years; e.g.][]{Burkhardt2019}, adding time constraints to the molecular compositions. 

Sites of shock chemistry include protostellar outflows that interact with the ambient medium. The outflow driven by the Class\,0 protostar L1157-mm (L1157 hereafter) has been the target of numerous projects, in which the source presented itself as a perfect laboratory to study COM chemistry \citep[][]{Bachiller1997}.
Single-dish and interferometric studies that reported detections of (complex) organic molecules in this source include, for example, \citet[][HCOOH, \mf, \metc]{Arce2008}, \citet[][\metc]{Codella2009}, \citet[][HCOOH, \mf, \ad]{Sugimura2011}, \citet[][HNCO, HCNO, \fmm]{Mendoza2014}, \citet[][HDCO, CH\2DOH]{Fontani2014}, \citet[][HCOOH, H\2CCO, \dme, HCOCH\2OH, \et]{Lefloch2017}, \citet[][HC\3N, HC\5N]{Mendoza2018}, and \citet[][\met, \ad]{Codella2020}.
Several of these studies concluded that COMs can be as abundant or even more abundant in shocked regions than in hot corinos \citep[e.g.][]{Arce2008,Mendoza2014,Lefloch2017} or shocked molecular clouds located in the Galactic centre \citep[][]{Arce2008}, such as G0.693$+$0.027 \citep[e.g. see Fig.\,11 in][]{Busch2024}.
In addition, spatial variations of COM emission provided insights into the likely dominating formation pathways. For example, \fmm emission peaks farther away from the protostar in the outflow than \ad \citep{Codella2017,Lopez-Sepulcre2024}, which has been interpreted as \fmm being formed later than \ad. Based on astrochemical models, the authors concluded that \fmm is dominantly formed in the post-shock gas phase, while \ad could be formed in the gas phase directly or desorbed from dust grain surfaces. 
With the knowledge gained from this template outflow, the search for other outflows is well underway. 
Complex organic molecules other than \met have been mapped towards the outflow system driven by the NGC1333 IRAS\,4A protobinary, which revealed emission from \ad, \dme, and \fmm \citep[][]{DeSimone2020}, and towards the HOPS\,409 outflow located in the OMC--2/3 filament \citep[\metc;][]{Bouvier2025}.

In this article, we present another young protostellar outflow and discuss its potential as new laboratory for shock chemistry. The outflow driven by the Class 0 protostar NGC1333 IRAS\,4B1, which forms a binary system with IRAS\,4B2 
\citep[2450\,au separation;][]{Tobin2016}, has been mapped in emission of various molecules, such as HCN \citep[][]{Choi2001}, H\2O \citep[][]{Desmurs2009}, H\2CO \citep[][]{diFrancesco2001}, H\2S and OCS \citep[][]{Miranzo-Pastor2025}, as well as CO, SiO, and SO \citep[][]{Jorgensen2007,Stephens2018,Podio2021} but COMs, other than \met \citep{Jorgensen2007,Sakai2012}, remained undetected. 
Additionally, \textit{Herschel}-PACS spectra towards the southern lobe revealed one of the richest far-infrared spectra amongst young low-mass protostellar sources with numerous highly excited H\2O, OH, CO emission lines \citep{Herczeg2012}. Following-up on this, the outflow has also been observed by the James Webb space telescope (JWST) as part of the JWST Observations of Young protoStars \citep[JOYS;][]{ewine2025,Francis2025}. The southern lobe was mapped in emission of H\2 S(1), [Fe\,II] $^4F_{7/2}$\,--\,$^4F_{9/2}$, and the CO\2 15\,$\mu$m band. Additional spectra show intense emission from hot C\2H\2 (600\,K) and HCN.
We report first detections of the following COMs towards the IRAS\,4B1 outflow: \metc, \ad, and deuterated methanol (CH\2DOH). We identify spatial variations in the outflow emission and derive abundances that provide insights into the outflow structure and kinematics as well as the impact of shocks on the chemical composition.

\section{Observations}
\subsection{PRODIGE}\label{s:obs}

The IRAS\,4B binary system was observed as part of the PROtostars \& DIsks: Global Evolution (PRODIGE, PIs: P.\,Caselli and Th.\,Henning) large program. PRODIGE is a MPG/IRAM Observatory Program (MIOP, Project ID L19MB) and targeted 30 Class 0/I protostellar systems in the Perseus molecular cloud \citep[$D=294\,\pm\,17$\,pc;][]{Zucker2019} with the Northern Extended Millimetre Array (NOEMA) at 1\,mm.

The data were taken with the Band 3 receiver and using the PolyFix correlator tuned to a local-oscillator (LO) frequency of 226.5\,GHz. 
This setup covers a total bandwidth of 16\,GHz with a channel width of 2\,MHz ($\sim$2.6\,\kms), divided into four sidebands: lower outer (214.7--218.8\,GHz), lower inner (218.8--222.8\,GHz), upper inner (230.2--234.2\,GHz), and upper outer (234.2--238.3\,GHz).
Additional 39 windows at high spectral-resolution (62.5\,kHz or $\sim$0.09\,\kms channel width), each covering a 64\,MHz bandwidth, were placed within the 16\,GHz bandwidth. 
The observations probe spatial scales of approximately 300\,au (corresponding to the highest angular resolution of $\sim$1\arcsec) to 5000\,au at the distance of Perseus. 
The phase centre is at $(\alpha,\delta)_\mathrm{J2000}=(03^\mathrm{h}29^\mathrm{m}12\overset{s}{.}02,31^\circ13^\prime08\overset{\prime\prime}{.}03)$.
The PRODIGE observations of IRAS\,4B were conducted in January and April 2022 in array configurations C and D, covering baselines from 24\,m to 400\,m.

For the data calibration of the PRODIGE data we used the standard observatory pipeline within GILDAS/CLIC\footnote{\url{https://www.iram.fr/IRAMFR/GILDAS/}} package.
Continuum subtraction and data imaging were performed with the GILDAS/MAPPING package using the \texttt{uv\_baseline} and \texttt{clean} tasks, respectively. For the imaging of the continuum maps we used robust\,=\,1, to improve the angular resolution, and for the spectral line cubes we used natural weighting to minimise noise.
More details on data reduction and imaging can be found in \citet{Gieser2024}. 

Integrated intensity maps were created using the high spectral-resolution data for selected molecules except for CO and HNCO (see Sect.\,\ref{ss:morph}). Table\,\ref{tab:lines} summarises all transitions used for the various maps with their spectroscopic information. The average noise in these cubes is about 10\,mJy\,beam$^{-1}$ (62.5\,kHz channel width). For the spectral-line analysis, spectra were extracted towards two selected positions from the low spectral-resolution (2\,MHz) cubes, as more molecular lines are covered (see Sect.\,\ref{ss:analysis}). These have noise levels of about 2\,mJy\,beam$^{-1}$.

\subsection{ALPPS}

Additionally, we made use of data taken with the Atacama Large Millimetre Array (ALMA) as part of the ALMA Perseus Polarization Survey \citep[ALPPS;][]{Cortes2025} towards IRAS\,4B1 at an angular resolution of $0.37\arcsec\times0.30\arcsec$ ($PA=-20.6^\circ$). The spectral setup covers a bandwidth of 1.9\,GHz from 335.5\,GHz to 337.4\,GHz at 0.976\,MHz spectral resolution ($\sim$0.9\,\kms). The noise level measured in the cube is 22\,mJy\,beam$^{-1}$.
We use the data to produce maps of the outflow emission at high angular resolution. Additional transitions of molecules analysed with the PRODIGE data are added to the temperature and column density determinations. To combine both data sets, we needed to smooth the ALPPS data to the angular (1.1\arcsec\,$\times$\,0.8\arcsec) and 2\,MHz spectral resolution of the PRODIGE data using the \texttt{convolve\_to} python command and the \texttt{resample} command in CLASS, respectively.

\section{Results}\label{s:results}

First, we study the outflow emission morphology of the detected molecules in Sect.\,\ref{ss:morph} and highlight differences based on maps showing ratios of integrated intensities with respect to \met in Sect.\,\ref{ss:ratios}. Molecular rotational temperatures and column densities are derived in Sect.\,\ref{ss:analysis} towards two positions, one in each outflow lobe, and assuming local thermodynamic equilibrium (LTE). The derived molecular composition is subsequently compared to the one in L1157-B1 in Sect.\,\ref{s:discussion}.

\begin{figure}[t]
    \centering
    \includegraphics[width=\linewidth]{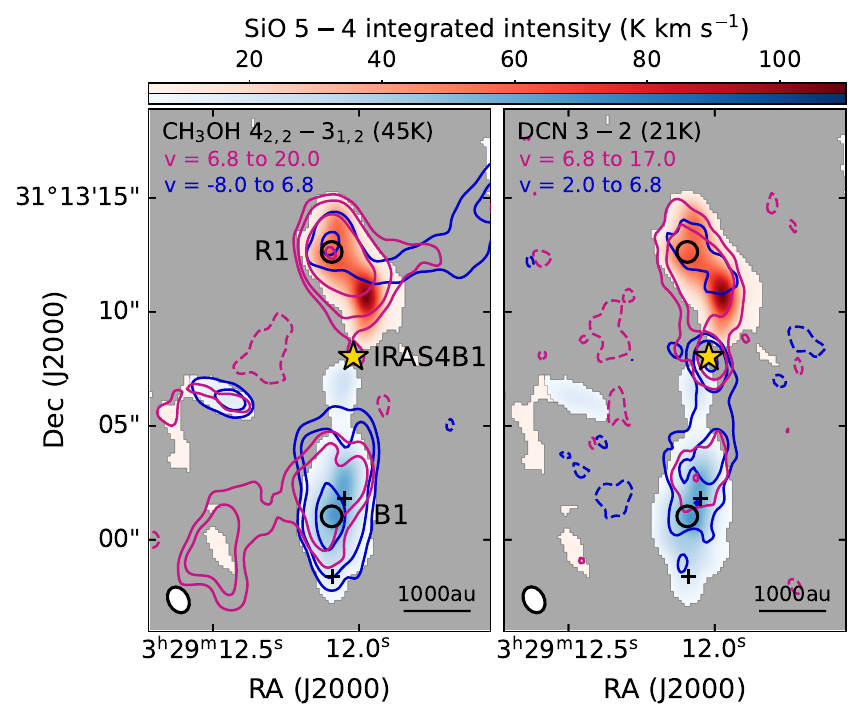}\vspace{-0.3cm}
    \caption{Integrated intensity maps towards IRAS\,4B1 (yellow star) for SiO 2\,--\,1 ($E_u=31$\,K; colour) and \met $4_2-3_1$ and DCN 3\,--\,2 ($E_u=45$\,K and 21\,K, respectively; contours). The SiO maps are integrated from $-25$ to 6\,\kms and 6.8 to 47\,\kms and show emission above 10$\sigma$ with $\sigma_\mathrm{blue}=0.42$\,K\,\kms and $\sigma_\mathrm{red}=0.49$\,K\,\kms. Contours are at $-$30$\sigma$,$-$15$\sigma$, 15$\sigma$, 30$\sigma$, and then increase by a factor of 3 for \met ($\sigma_\mathrm{blue}=0.25$\,K\,\kms and $\sigma_\mathrm{red}=0.23$\,K\,\kms) and at $-$10$\sigma$, $-$5$\sigma$, 5$\sigma$, 10$\sigma$, and then increase by a factor of 3 for DCN ($\sigma_\mathrm{blue}=0.17$\,K\,\kms and $\sigma_\mathrm{red}=0.24$\,K\,\kms). Velocity ranges (in \kms) used for the integration of \met and DCN emission are indicated in the top left. The HPBW is shown in the bottom left. Positions R1 and B1 were selected for further spectral-line analysis (Sect.\,\ref{ss:analysis}). Black crosses indicate H\2 knots \citep[][]{Choi2011}.}
    \label{fig:main}
\end{figure}
\begin{figure}[t]%
    \includegraphics[width=\linewidth]{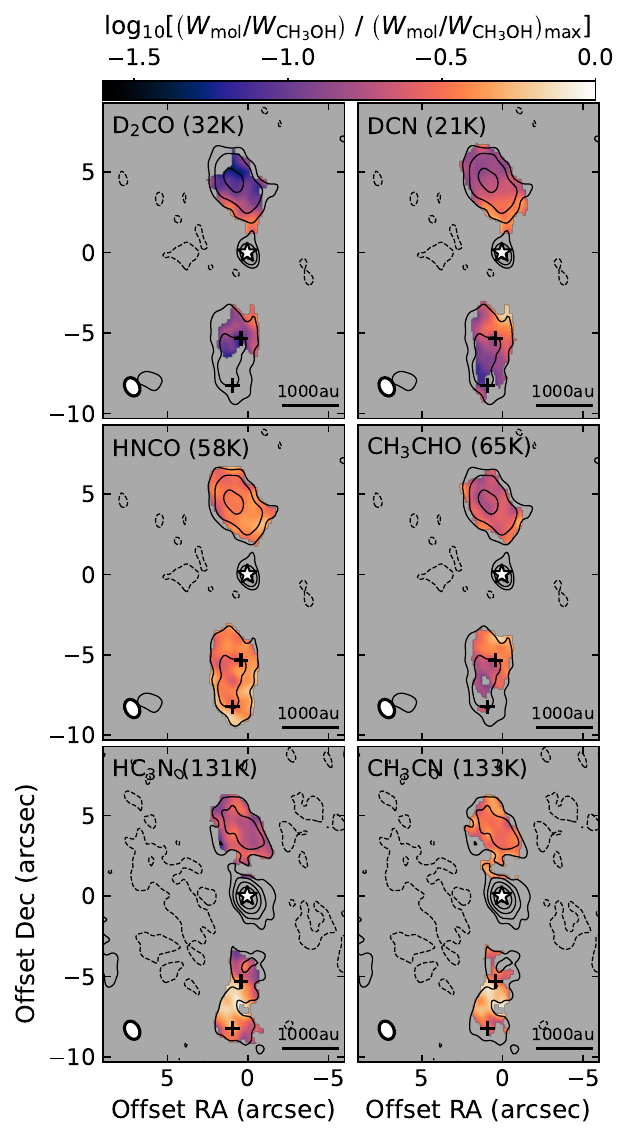}\vspace{-0.3cm}
    \caption{Integrated intensity ratios, $(W_\mathrm{mol} / W_\mathrm{CH_3OH})$ / $( W_\mathrm{mol} / W_\mathrm{CH_3OH})_\mathrm{max}$ of various molecular transitions with either \met $10_{2,9}-9_{3,6}$ ($E_u=165$\,K; for HC\3N and \metc) or $5_{1,4}-4_{2,2}$ ($E_u=56$\,K; for D\2CO, DCN, HNCO, and \ad) towards the IRAS\,4B1 outflow. Integrated intensities contain the sum of blue- and redshifted emission (cf. Fig.\,\ref{fig:mom0} and spectra in Fig.\,\ref{fig:spectra}). The ratio is only shown if both molecules are above a 5$\sigma$ threshold, where $\sigma$ is the noise level in the respective map. Another mask with a 1\arcsec\,\,radius around the protostar is applied to the ratios. 
    Black contours show the integrated intensities of \met, starting at $-5\sigma$, 5$\sigma$, and then increasing by a factor of 2, where $\sigma=0.34$\,K\,\kms. The molecule as well as the upper-level energy of the shown transition are in the top left corner. In all panels, the white star marks the position of IRAS\,4B1. The HPBW is shown in the bottom left corner and black crosses mark H\2 knots \citep{Choi2011}. Spectroscopic information on the shown transitions are given in Table\,\ref{tab:lines}.}
    \label{fig:ratios}
\end{figure}

\subsection{Emission morphology}\label{ss:morph}

Figure\,\ref{fig:main} shows integrated intensity maps towards IRAS\,4B1 of SiO 2\,--\,1 (colour-scales), \met $4_2-3_1$, and DCN 3\,--\,2 (contours), separated in red- and blueshifted emission \citep[with respect to the systemic velocity $\varv_\mathrm{sys}=6.8$\,\kms;][]{Busch2025}.
The emission extends to the north and south from the central protostar and can be associated with the outflow driven by that protostar. 
Red- and blueshifted emission at velocities of $\Delta\varv=|\varv_\mathrm{sys}-\varv|<10$\,\kms  is present in both lobes, suggesting that the outflow lies close to the plane of the sky. The overlap of red- and blueshifted emission is also observed for SiO (cf. Figs.\,\ref{fig:spectra} and \ref{fig:chan_maps}), however, in Fig.\,\ref{fig:main}, we only show emission  above 10$\sigma$ to focus on the dominant SiO features in each lobe. 
Emission from \met and DCN spatially coincides with SiO for the most part. Blueshifted SiO emission close to the protostar in the southern lobe and a peak in redshifted SiO emission at $\sim$(0\arcsec, 2.5\arcsec) are not evident in \met emission, while DCN does not extend as far as SiO and \met in the southern lobe.
In addition, \met reveals arc-like emission features that seem to start from within the main outflow lobes in the north and south and continue NW and SE, respectively. While the NW extension contains red- and blueshifted emission, the SE extension only appears in redshifted emission in our data.
These features were proposed to be a second outflow, 
based on larger-scale methanol maps, which extend beyond our field of view \citep{Sakai2012}. The authors discussed the possibility of IRAS\,4B1 consisting of two protostars, each driving an outflow.
However, the presence of a second source within IRAS\,4B1 remains debated. At spatial scales of $\lesssim$\,30\,au, continuum maps at 8\,mm show no clear evidence for a companion \citep[see Fig.\,34 in][]{Segura-Cox2018}, although there is extended fainter emission in addition to the peak. 
Emission to the east in Fig.\,\ref{fig:main} originates from the outflow driven by the companion IRAS\,4B2 (see also Figs.\,\ref{fig:chan_maps} and \ref{fig:mom0}) that is not discussed here further.
Low-excitation molecular transitions observed in the 3\,mm window reveal an overall complex kinematic structure, which will be studied in detail in a forthcoming paper. Such structures include, for example, the arcs identified here and a streamer candidate \citep[][]{Validvia-Mena2024}. 

Figure\,\ref{fig:chan_maps} shows additional channel maps of the typical outflow tracers SiO and CO, where intensities were integrated over 10\,\kms intervals to study substructure at different velocities.
At high blueshifted ($-$30 to $-$5\,\kms) and redshifted (20 to 45\,\kms) velocities, the emission of the IRAS\,4B1 outflow is confined within the southern and northern lobes, respectively, in contrast to the broader lower-velocity emission. This can result from the fact that this higher-velocity component is more inclined or more collimated than the lower-velocity component. The high-velocity emission can likely be associated with a jet component \citep[see also][]{Podio2021}.
The presence of various intensity peaks or knots along both lobes may additionally indicate episodic ejection \citep[e.g.][]{Bachiller1996}.  

Figure\,\ref{fig:mom0} shows integrated intensity maps of blue- and redshifted emission again for \met and DCN, and for seven additional organic (i.e. C-bearing) molecules. Example spectra towards positions R1 and B1 (see Fig.\,\ref{fig:main}) are shown in Fig.\,\ref{fig:spectra}.
Emission from H\2CO and \met in Fig.\,\ref{fig:mom0} is the most prominent in the outflow. Molecules such as \ad, DCN, HC\3N, \metc, HNCO, and D\2CO show weaker emission. For HC\3N and \metc, but maybe also HNCO and \ad, this could be an excitation effect, since the PRODIGE data only cover transitions with higher upper-level energies. Weak (signal-to-noise, S/N $\sim$\,5), compact emission from $c$-C\3H\2 is detected very close to the protostar. 
Besides \met, H\2CO and CO reveal the additional arc-like emission features (Figs.\,\ref{fig:chan_maps} and \ref{fig:mom0}).
There are extended features of negative intensity for, in particular, CO and H\2CO, which point to missing flux in the PRODIGE data due to missing short spacings \citep[see also][]{Gieser2025}.

\subsection{Spatial differences with respect to \met}\label{ss:ratios}

To see whether there are morphological differences, we compare all molecules with \met by deriving ratio maps of integrated intensity, $W$, that is $R = W_\mathrm{mol}/W_\mathrm{CH_3OH}$, and show them normalised by the maximum value, $R/R_\mathrm{max}$, in Fig.\,\ref{fig:ratios}. We only plot pixels, where both integrated intensity values are above a 5$\sigma$ threshold, where $\sigma$ is the noise measured in the respective map. The integration intervals differ between molecules (same intervals as in Figs.\,\ref{fig:spectra} and \ref{fig:mom0}). If we used the same (largest) interval, we would integrate over too much noise leading to more discarded pixels for the weakest transitions. In addition, we apply a mask of a 1\arcsec\,\,radius around the protostar because line blending is more severe in the hot corino and might bias the integrated intensity ratios. 
Because upper-level energies differ between the molecular transitions, meaning that they may probe different excitation conditions, we used two methanol transitions: $10_2-9_3$ ($E_u=165$\,K) with HC\3N and \metc ($E_u\approx130$\,K) and $5_1-4_2$ ($E_u=56$\,K) with the other molecules ($E_u=20-65$\,K), to reduce excitation effects. Additional information on the used transitions are summarised in Table\,\ref{tab:lines}.
There is a clear decrease in D\2CO and DCN intensities relative to \met with increasing distance from the protostar in both lobes by a factor of $\sim$10. 
In the southern lobe, a similar but less distinct gradient is seen for \ad.
For HC\3N and \metc, the integrated intensity ratio peaks in between the two H\2 knots \citep[][]{Choi2011} in the southern lobe. 
These trends may still (partially) be a consequence of different excitation conditions but may also probe different chemical processes. This is discussed further in Sect.\,\ref{s:discussion}.

\subsection{Spectral-line analysis \& results}\label{ss:analysis}

\begin{figure}[t]
    \centering
    \includegraphics[width=\linewidth]{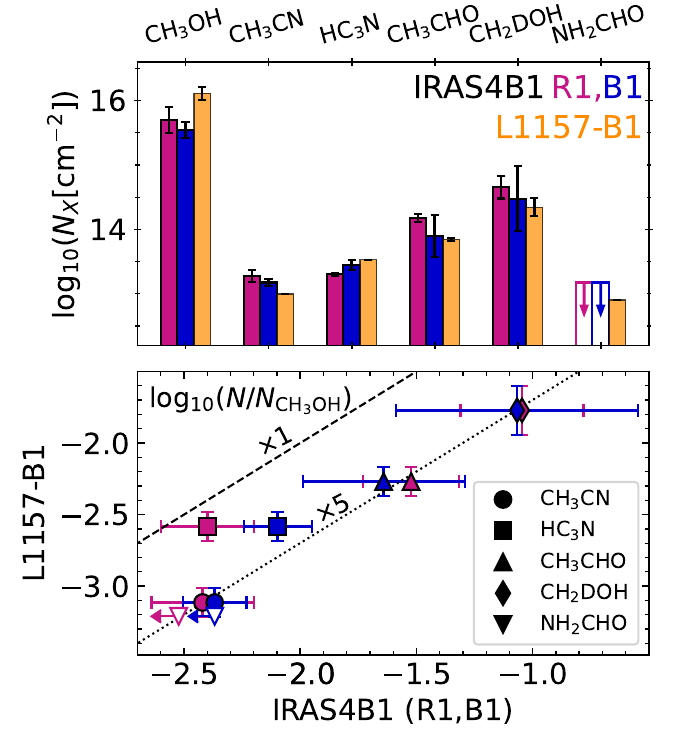}\vspace{-0.3cm}
    \caption{Column densities \textit{(top)} and abundances with respect to \met \textit{(bottom)} towards B1 and R1 in IRAS\,4B1 derived in this work and values obtained for L1157-B1 (see Table\,\ref{tab:analysis}). The dashed and dotted black lines indicate equal abundances and a difference of a factor of 5, respectively. Empty bars or symbols with arrows indicate upper limits.}
    \label{fig:l1157}
\end{figure} 

For the spectral-line analysis, we extracted beam-averaged spectra towards one position in the southern lobe (B1) and one in the northern lobe (R1) at $(03^\mathrm{h}29^\mathrm{m}12\overset{s}{.}09,31^\circ13^\prime01\overset{\prime\prime}{.}03)$ and $(03^\mathrm{h}29^\mathrm{m}12\overset{s}{.}09,31^\circ13^\prime12\overset{\prime\prime}{.}63)$, respectively, which correspond to the \met emission peaks (Fig.\,\ref{fig:main}). 
We follow the analysis strategy of \citet{Busch2025}, where we already derived rotational temperatures and column densities for \met and \metc isotopologues towards the hot corino IRAS\,4B1 amongst other sources. Accordingly, we use the radiative transfer code Weeds, which is an extension of the GILDAS/CLASS software \citep{Maret2011}, and derive population diagrams \citep[PDs;][]{Goldsmith1999}. Both methods assume local thermodynamic equilibrium (LTE). 
We refer to \cite{Busch2025} for details; in short: Weeds computes synthetic spectra based on five input parameters, which are column density, rotational temperature, source size, velocity offset from $\varv_\mathrm{sys}$, and line width (i.e. full width half maximum). 
We assume that the emission is spatially resolved (i.e. beam filling factor $=$ 1). 
Line widths and velocity offsets can be derived from fitting 1D Gaussian profiles to the spectral lines, using the CLASS command \texttt{minimize}, where we fitted one velocity component per molecule. 

The best-fit column densities and rotational temperatures are verified with values from PDs, which are shown in Fig.\,\ref{fig:PD}. 
The PDs were derived for \met, \ad, \metc, HC\3N, CH\2DOH, and D\2CO, since they are the only molecules with at least three detected transitions. 
The temperatures and column densities derived with Weeds and the PDs are summarised in Table\,\ref{tab:analysis} together with the other Weeds parameters. 
Results from Weeds and the PDs agree widely within the 1$\sigma$ error bars. Some scatter between the observed data points is evident (e.g. D\2CO in B1). We discussed several reasons leading to such scatter when analysing the PRODIGE data in \citet[][]{Busch2025}. Weeds estimates line optical depths based on the input parameters, yielding optically thin lines in all cases. In some PDs, when the linear fit was unreliable, we fixed the rotational temperature to get an estimate on the column density.
%
Rotational temperatures range from 50\,K to 100\,K without significant differences between positions B1 and R1.
Column densities derived towards B1 and R1 agree within a factor of 2 and are shown in Fig.\,\ref{fig:l1157}. 
They are discussed in a direct comparison with values derived towards the L1157 outflow in Sect.\,\ref{dss:compare}.

\section{Discussion: A new shock-chemistry laboratory}\label{s:discussion}

\begin{figure}
    \hspace{-0.3cm}
    \centering\includegraphics[width=\linewidth]{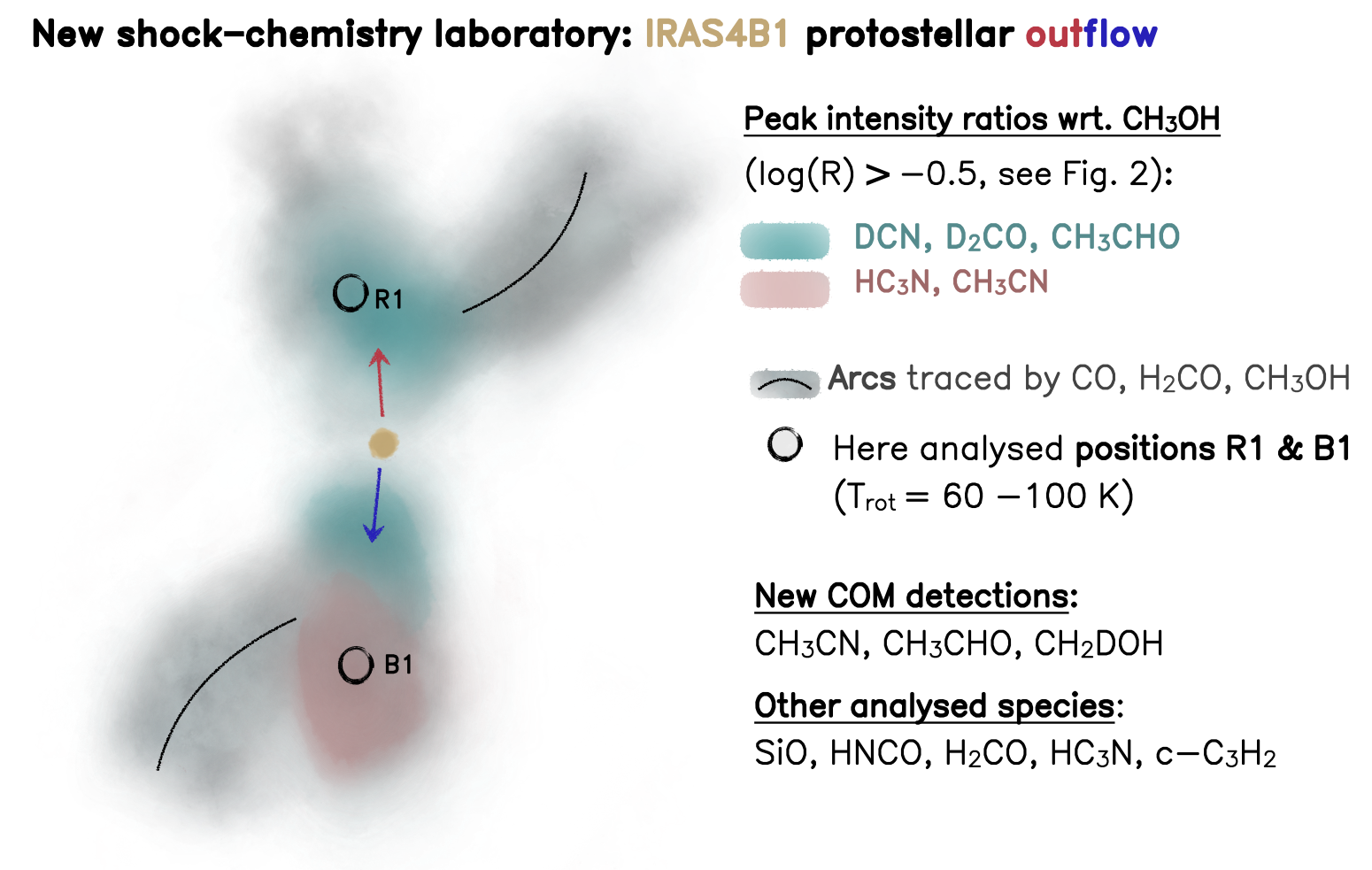}
    \caption{Cartoon highlighting the main molecular features of the IRAS\,4B1 outflow analysed in this work. The main emission morphology of our reference molecule, \met, is outlined in grey, including the main outflow lobes (red and blue arrows) and additional arc-like features (black lines) of currently unknown origin. Peaks of intensity ratios with respect to \met are highlighted in teal and pink, depending on the species (cf. Fig.\,\ref{fig:ratios}). A spectral-line analysis was done towards positions R1 and B1 that yielded rotational temperatures of $T_\mathrm{rot}=60-100$\,K (Sect.\,\ref{ss:analysis}). The position of the protostar is indicated in yellow.}
    \label{fig:cartoon}
\end{figure}

Figure\,\ref{fig:cartoon} summarises the main molecular features of the IRAS\,4B1 outflow derived in this work, including the morphological differences of molecules with respect to \met, which are further discussed in Sects.\,\ref{dss:deuterated} and \ref{dss:N}. In addition, we looked at the morphology at higher angular resolution using ALPPS data (Sect.\,\ref{dss:alpps}). In Sect.\,\ref{dss:compare} we compare our results with the ones obtained for the L1157 outflow. 

\subsection{Implications from the morphology}

Differences in the emission morphology between molecules can probe different physical conditions but may also be explained by certain chemical processes.
In Fig.\,\ref{fig:ratios} we mapped the outflow emission from different species in comparison with \met. Figure\,\ref{fig:cartoon} highlights the main findings. Since \met is known to be formed on dust-grain surfaces and is released into the gas phase due to the shocks, it is a suitable reference molecule to interpret any morphological differences with other molecules. 

\subsubsection{Deuterated species}\label{dss:deuterated}

The intensity ratios for the deuterated species DCN and D\2CO with \met (Fig.\,\ref{fig:ratios} and \ref{fig:cartoon}) show a pronounced decrease with increasing distance from the protostar in both lobes. Recently, D\2CO emission was studied towards the IRAS\,4A outflows \citep{Chahine2024}, where it shows a similar trend.
During the prestellar stage, D\2CO is more efficiently formed on dust-grain surfaces the lower temperatures are, that is closer to the centre of the prestellar core. Later, upon interaction with the outflow, D\2CO is then released from the grain surfaces. Therefore, the observed gradient along the outflow lobes may reflect the abundance profile that was built up during the prestellar phase, provided the molecules were recently released from grains and gas-phase chemistry has not altered the abundance distribution yet \citep[see also][]{Podio2024}. 
Whether this also applies to DCN is not entirely clear. Based on a study of DCN towards the L1157 outflow, \citet{Busquet2017} argue that the molecule is a product of both gas- and solid-phase chemistry. 
Moreover, it is not clear whether DCN or HCN freeze-out as efficiently during the prestellar phase. It was shown in prestellar cores that HCN remains in the gas phase where CO is already frozen-out \citep[e.g.][]{Hily-Blant2010,Spezzano2022}. 
These single-dish studies probed larger spatial scales, meaning that freeze-out of DCN and HCN at smaller scales is not excluded when probed with interferometers. Model predictions are also inconclusive as they are heavily dependent on the underlying physical setup (priv. comm.). 

\subsubsection{\ad, HC\3N, and \metc}\label{dss:N}

\begin{figure}[t]
    \centering
    \includegraphics[width=\linewidth]{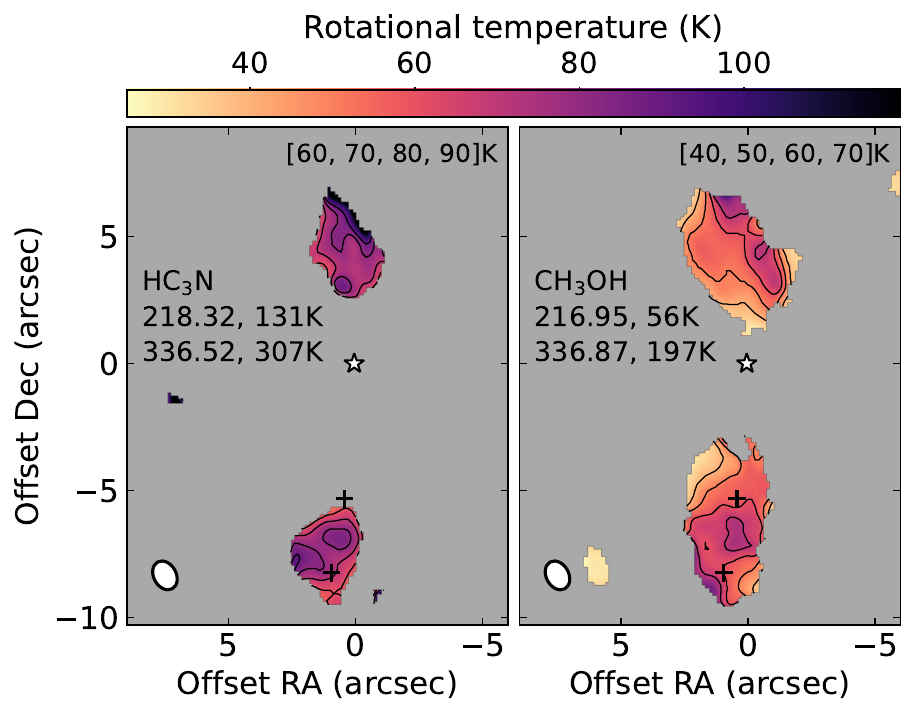}\vspace{-0.3cm}
    \caption{Rotational temperatures derived from two transitions of HC\3N and \met assuming LTE. Rest frequencies (in GHz) and upper-level energies of the transitions are written below the molecules. Contour levels are shown in the top right. Markers are the same as in Fig.\,\ref{fig:ratios}. The HPBW is shown in the bottom left corner.}
    \label{fig:temp}
\end{figure}

In the southern outflow lobe of IRAS\,4B1, we observed decreasing \ad intensity ratios relative to \met (Fig.\,\ref{fig:ratios}) with increasing distance from the protostar, while emission from both COMs is more similar in the northern lobe.
In L1157-B1, \ad and \met coincide spatially \citep[][]{Codella2020}, implying that \ad is also formed in the solid phase or, as proposed by \citet{Codella2017,Codella2020}, rapidly in the gas phase as soon as the reactants arrive in the gas phase from the grains. The gradient that we observe may suggest that abundances drop faster for \ad than for \met starting from IRAS\,4B1. In contrast to \met, \ad can also efficiently form in high-temperature gas \citep[e.g.][]{Garrod2022}.
Therefore, to see whether the observed emission gradient coincides with a temperature gradient, we derived maps of rotational temperature using two transitions with different upper-level energies and assuming LTE and that the transitions trace the same gas. The maps for HC\3N and \met are shown in Fig.\,\ref{fig:temp} and reveal temperatures from 30--100\,K with HC\3N probing overall slightly higher temperatures. However, the general trends are similar for both molecules: in the northern lobe, temperature increases from east to west within the northern lobe. In the southern lobe, temperature increases with increasing distance from the protostar, peaks in between the two H\2 knots, and decreases again at the tip of the lobe.
The observed emission gradient between \ad and \met is thus anticorrelated with the temperature gradient in the southern lobe. It is not clear to us, what this may infer. Instead of temperature, other parameters, such as density variations or cosmic rays \citep[][]{Pineda2024} may play a more dominating role. 

The intensity ratios for HC\3N and \metc (Fig.\,\ref{fig:ratios} and \ref{fig:cartoon}) follow the temperature map, more in the southern than the northern lobe, suggesting that their abundances are enhanced in the higher temperature gas. Gas-phase formation was also proposed for \metc and HC\3N in the L1157 outflow \citep[][]{Codella2009,Mendoza2018}. For \metc, this may also be supported by the fact that the column density ratio between \met and \metc, which is $\sim$0.005 towards the IRAS\,4B1 outflow, is a factor of 2 smaller than the one derived towards the central hot corino \citep[][]{Busch2025}, where even higher temperatures can facilitate gas-phase formation of \metc \citep[][]{Giani2023}. In addition, JWST spectra revealed rich spectra including features from HCN and C\2H\2 \citep[][]{ewine2025} that may be involved in the enhanced production of HC\3N and \metc \citep[e.g.][]{Taniguchi2019}.

\subsubsection{Implications from ALPPS data}\label{dss:alpps}

Observations of the L1157 outflow at high angular resolution revealed that the L1157-B1 shock spot in the blueshifted lobe consists of several smaller clumps, whose molecular compositions and physical conditions differ \citep[][]{Benedettini2007,Benedettini2013,Codella2009,Codella2020}, offering insights into the outflow structure, kinematics, and chemistry. 
Using ALPPS data, we show integrated intensity maps of \met $7_{1,7}-6_{1,6}$ (the only bright COM covered in that survey) and C$^{34}$S 7\,--\,6 in Fig.\,\ref{fig:chan_maps}. Similar to L1157-B1, these observations at $\sim$3$\times$ higher angular resolution than PRODIGE reveal a highly structured outflow morphology with several intensity peaks. In addition, the morphology of the \met emission in the southern lobe resembles a bow-shock, while the C$^{34}$S emission forms a weak S-shape, which is indicative of jet precession \citep[e.g.][]{Bachiller2001}. Therefore, future observations of more species at higher angular resolution will provide deeper insight into the outflow's kinematics and chemistry.

\subsection{Comparison with L1157-B1}\label{dss:compare}

We compare the column densities and abundances with respect to \met that we derived towards positions IRAS\,4B1-B1 and R1 with literature values from L1157-B1, which were derived in interferometric observations, in Fig.\,\ref{fig:l1157} (see Table\,\ref{tab:analysis}). 
Overall, column densities agree within a factor of 2 between the sources, while abundances with respect to \met are higher towards R1 and B1 in IRAS\,4B1. Most abundances differ by a factor 5, suggesting that the chemistry may essentially be the same. One of the reasons for the difference of a factor 5 may be an underestimation of optical depth, hence column density of \met, which is quite abundant in the outflow.  
The overall similar chemical composition is interesting given that both outflow showcase different physical conditions. 
We derived the molecular composition at positions that are much closer to IRAS\,4B1 ($D\sim1500-2000$\,au) than L1157-B1 is to its protostar \citep[$D\sim30\,000$\,au;][]{Arce2008}.
Related to that, rotational temperatures that we derived, for example, for \met and \ad (60--80\,K) are significantly higher than towards L1157-B1 but also the IRAS\,4A outflow \citep[$\sim$10--20\,K;][]{DeSimone2020,Codella2020}. On the other hand, rotational temperatures derived from \metc in L1157-B1 range from 50\,K to 130\,K \citep{Codella2009}, similar to what we derived for this molecule. 
Generally, higher temperatures suggest that volume densities are higher, which may hint at stronger shocks \citep[][]{McKee1980}.
Moreover, the observed subthermal excitation of H\2O at 1500\,K observed in \textit{Herschel}-PACS spectra in the southern lobe requires a density  of the order of 10$^{6}$\,cm$^{-3}$ \citep{Herczeg2012}, which makes the IRAS\,4B1 outflow a site of a high (pre-)shock density \citep[see also][]{ewine2025}. 

The different maximum spatial extent also implies different outflow ages or shock timescales. The compact IRAS\,4B1 outflow was proposed to be very young \citep[a few 100\,yr;][]{Choi2001,Yildiz2015}, and is, therefore, younger than the L1157 outflow \citep[a few 1000\,yr; e.g.][]{Yildiz2015,Kwon2015,Podio2016}. 
The timescale associated with the age of the IRAS\,4B1 outflow could be crucial for post-shock gas-phase chemistry. For example, a period of $\lesssim$1000\,yr is associated with the delayed formation of \fmm compared with \ad in L1157-B1, which was mentioned in the introduction \citep[Sect.\,\ref{s:intro};][]{Codella2017}. We do not detect \fmm. If it remains undetected in more sensitive future observations towards the outflow of IRAS\,4B1, this may suggest that the chemistry in the post-shock gas is in an early phase, meaning that a significant amount of \fmm has not yet formed. Therefore, the young age of the IRAS\,4B1 outflow may be the perfect test bed to study grain-surface versus gas-phase formation of COMs.

\section{Summary}\label{s:conclusion}

We used data taken as part of the PRODIGE large program and ALPPS to study the emission of organic molecules, including COMs, towards the outflow driven by the Class 0 protostar NGC1333 IRAS\,4B1. We posit the potential of this outflow to be a new laboratory for the study of COMs in shocked media.  Our main findings are the following:
\begin{itemize}
    \item We report the first detection of the COMs \metc, \ad, and deuterated methanol (CH\2DOH) towards this outflow. 
    \item We investigated the morphology of the outflow emission of these three COMs, \met, and simpler species (HC\3N, HNCO, CO, DCN, D\2CO, H\2CO, $c-$C\3H\2). 
    The deuterated species DCN and D\2CO showed a gradient with respect to \met, where they would peak closer to the protostar. This likely reflects the DCN and D\2CO abundance profiles from the prestellar phase. Acetaldehyde (\ad) showed a similar gradient in the southern lobe, which still needs an explanation.
    In the southern lobe, intensity ratios of HC\3N and \metc with respect to \met followed the temperature variations, suggesting that the N-bearing molecules are (additionally) formed in the hotter post-shock gas phase. 
    \item Methanol emission at high angular resolution using ALPPS data, reveal a highly structured outflow morphology, which is also seen in L1157-B1.
    \item Towards one position in the northern lobe (R1) and one in the southern lobe (B1), we derived on average higher rotational temperatures (50--100\,K) and higher abundances with respect to \met by factors of a few than towards L1157-B1.
\end{itemize}
Future observations at higher angular resolution that are as sensitive as those towards L1157-B1 will enable us to detect less abundant COMs and build a comprehensive chemical inventory of the IRAS\,4B1 outflow, which we can compare to the one of the central hot corino. Together with modelling efforts, this will deliver crucial information on COM formation and destruction processes as well as outflow structure and kinematics.

\begin{acknowledgements}
The authors thank the IRAM staff at the NOEMA observatory for their support in the observations and data calibration. This work is based on observations carried out under project number L19MB with the IRAM NOEMA Interferometer. IRAM is supported by INSU/CNRS (France), MPG (Germany) and IGN (Spain). L.A.B, J.E.P., P.C., M.J.M, C.G., Y.L., D.S., Y.C., L.M., S.N. acknowledge the support by the Max Planck Society. D.S. was funded by the Deutsche Forschungsgemeinschaft (DFG, German Research Foundation) – project number: 550639632. This project is co-funded by the European Union (ERC, SUL4LIFE, grant agreement No101096293). A.F. also thanks project PID2022-137980NB-I00 funded by the Spanish Ministry of Science and Innovation/State Agency of Research MCIN/AEI/10.13039/501100011033 and by “ERDF A way of making Europe”.
\end{acknowledgements}

\bibliographystyle{aa} 
\bibliography{refs} 


\appendix
\onecolumn

\section{Additional tables and figures}

\begin{figure}[h]
    \centering
    \includegraphics[width=.9\linewidth]{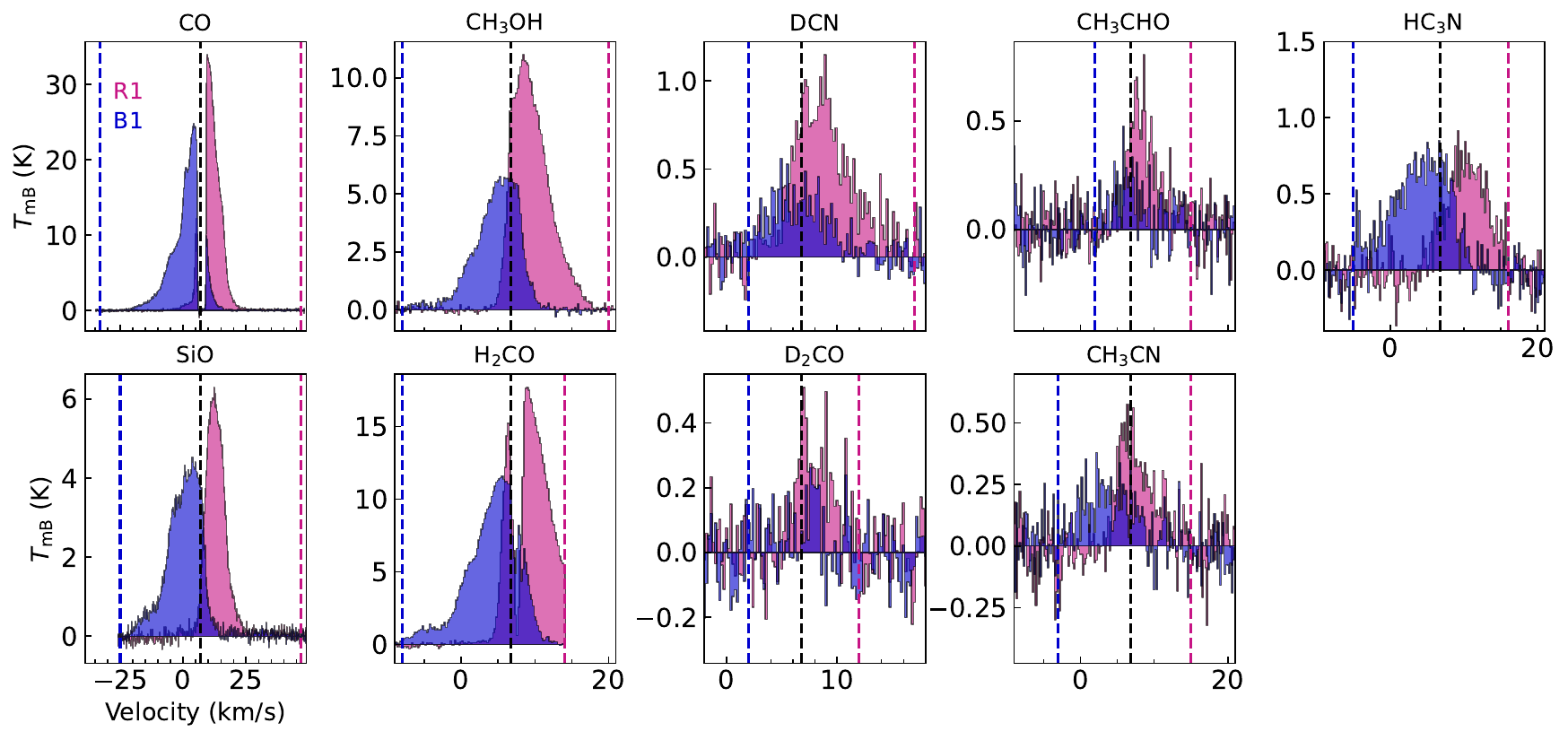}\vspace{-0.3cm}
    \caption{Average spectra of selected molecules extracted from a 1\arcsec\,\,aperture towards positions R1 and B1 (see Fig.\,\ref{fig:main}). The black dashed line indicates the systemic velocity of 6.8\,\kms, while red and blue dashed lines indicate the outer integration limits for the respective molecule used in Figs.\,\ref{fig:main}, \ref{fig:ratios}, and \ref{fig:mom0}. Despite the narrow band not covering all H\2CO emission, we used it for the H\2CO map as it shows more structure.} 
    \label{fig:spectra}
\end{figure}

\begin{figure*}[h]
    \centering
    \includegraphics[width=0.352\linewidth]{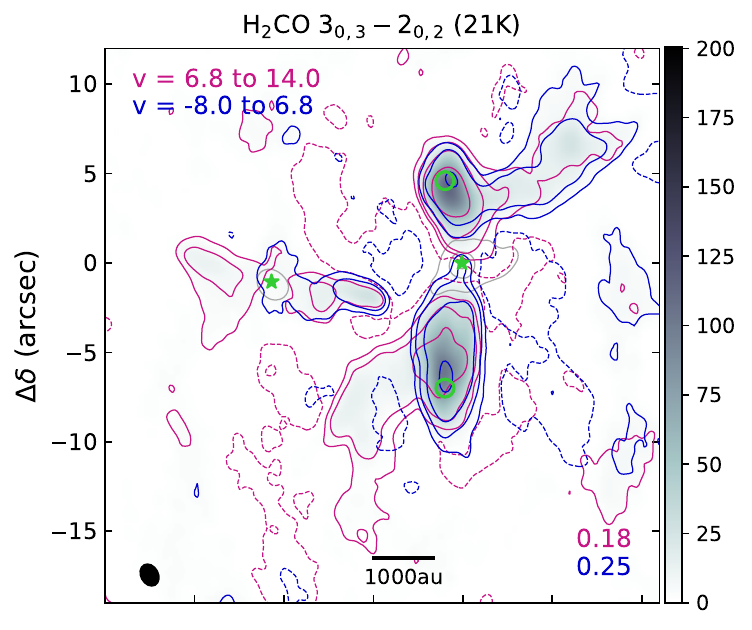}
    \includegraphics[width=0.299\linewidth]{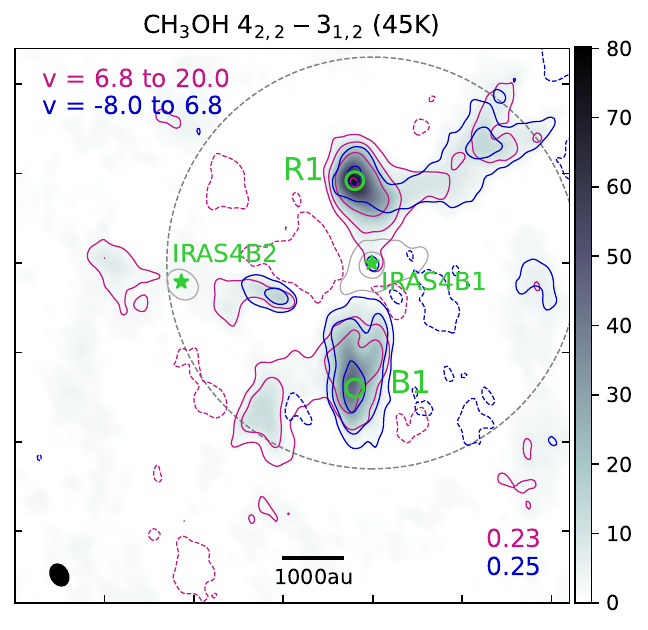}
    \includegraphics[width=0.309\linewidth]{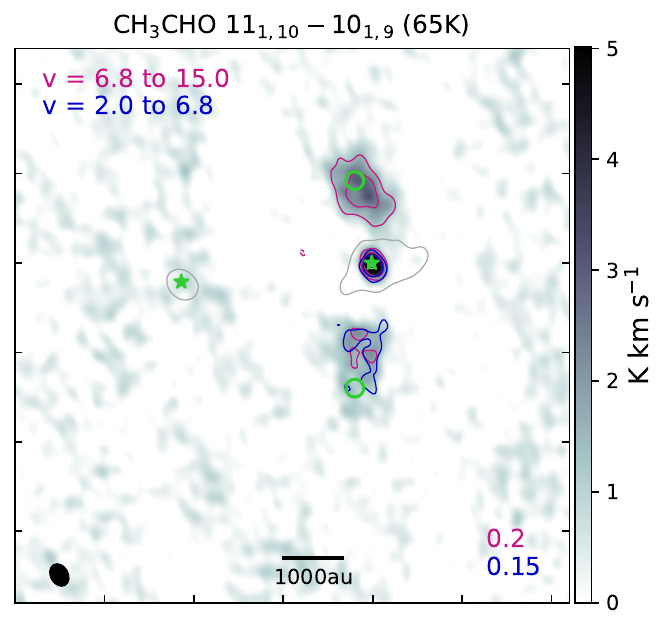}
    \includegraphics[width=0.338\linewidth]{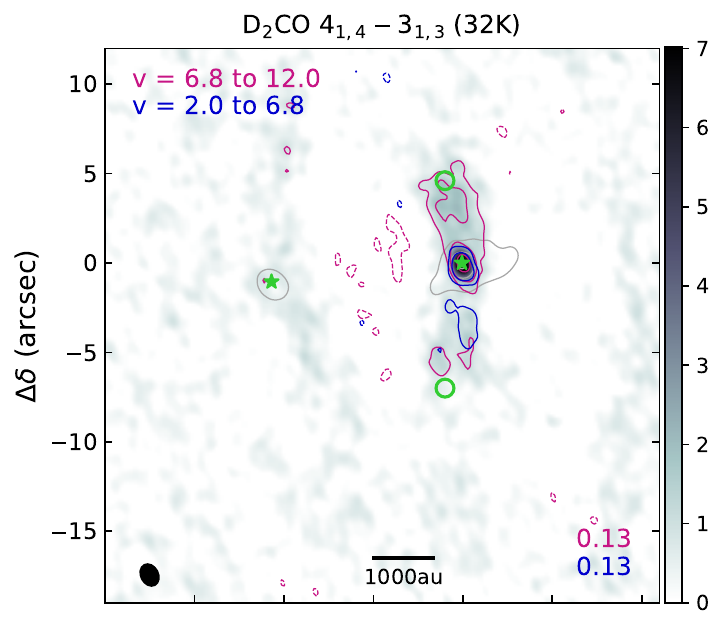}\hspace{0.3cm}
    \includegraphics[width=0.3\linewidth]{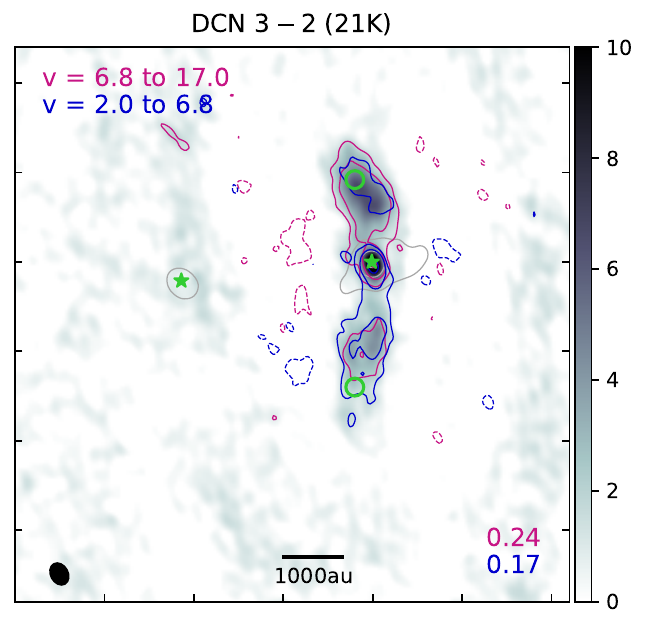}
    \includegraphics[width=0.32\linewidth]{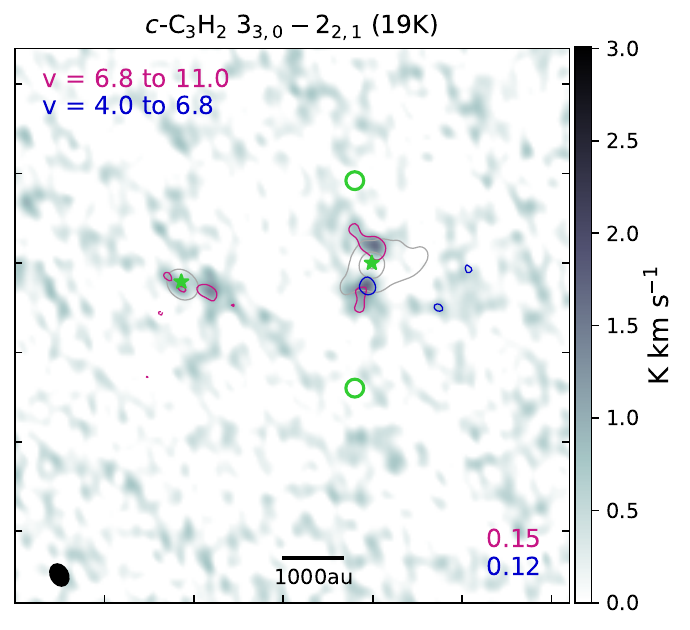}
    \includegraphics[width=0.342\linewidth]{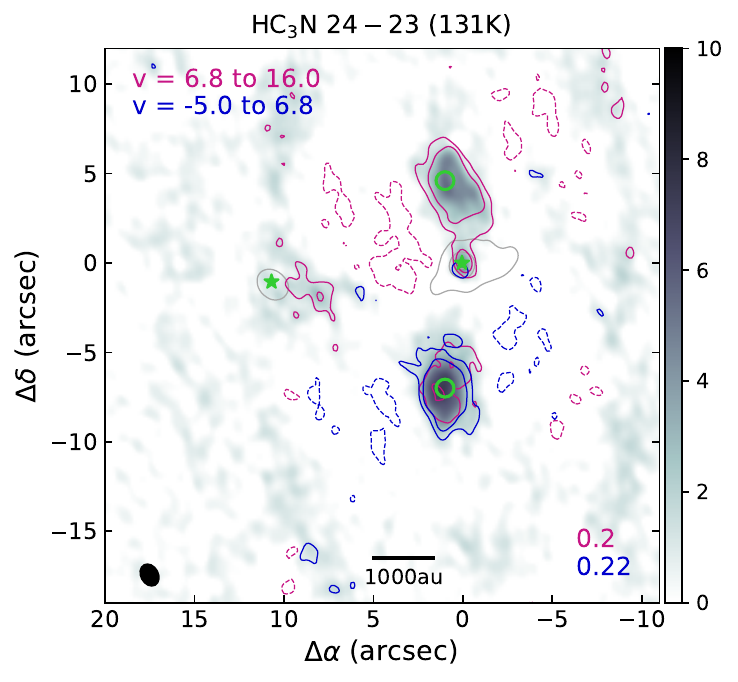}
    \includegraphics[width=0.297\linewidth]{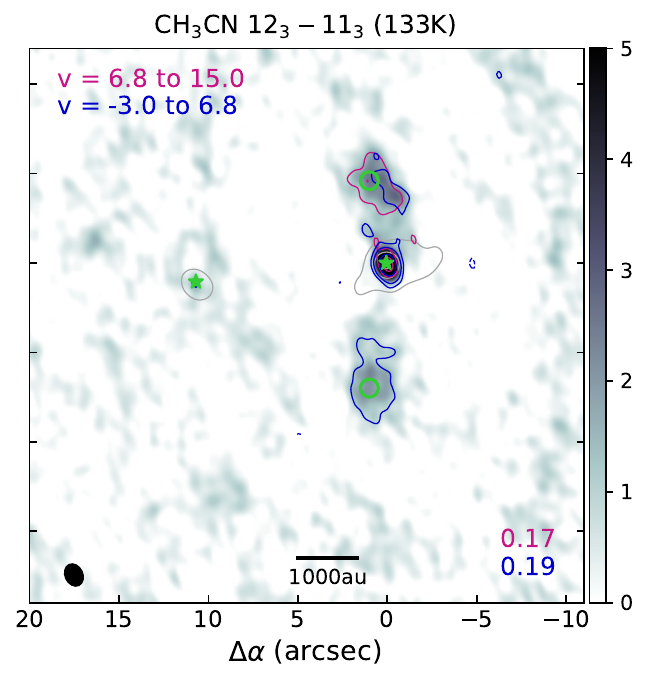}
    \includegraphics[width=0.322\linewidth]{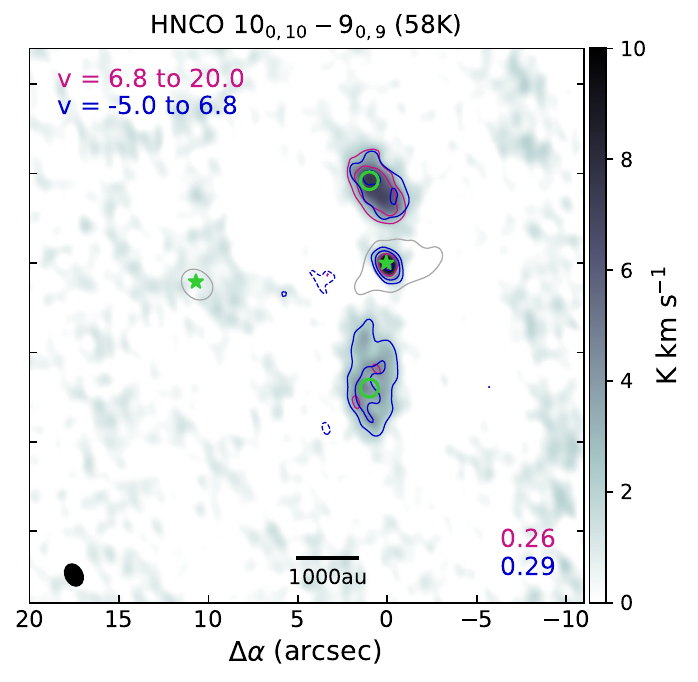}
    
    \caption{Integrated intensity maps towards IRAS\,4B1 and 4B2 (green stars) for the molecular transitions that are written on top with their respective upper-level energy. Contours are at $-$5$\sigma$ 5$\sigma$, 10$\sigma$, and then increase by a factor 3 for all but SiO, CO, H\2CO, and \met, for which contours start at $-$10$\sigma$ and 10$\sigma$. The noise levels, $\sigma$, are measured in the respective maps and are written in the bottom right. Velocity (in \kms) ranges for integration are indicated in the top left. The grey scale shows the sum of the red- and blueshifted maps. The HPBW is shown in the bottom left. The dashed grey circle indicates the primary beam ($\sim$23\arcsec). Positions R1 and B1 (green circles) are selected for further spectral-line analysis.} 
    \label{fig:mom0}
\end{figure*}

\begin{figure}[h]
    \centering
    \includegraphics[width=0.47\textwidth]{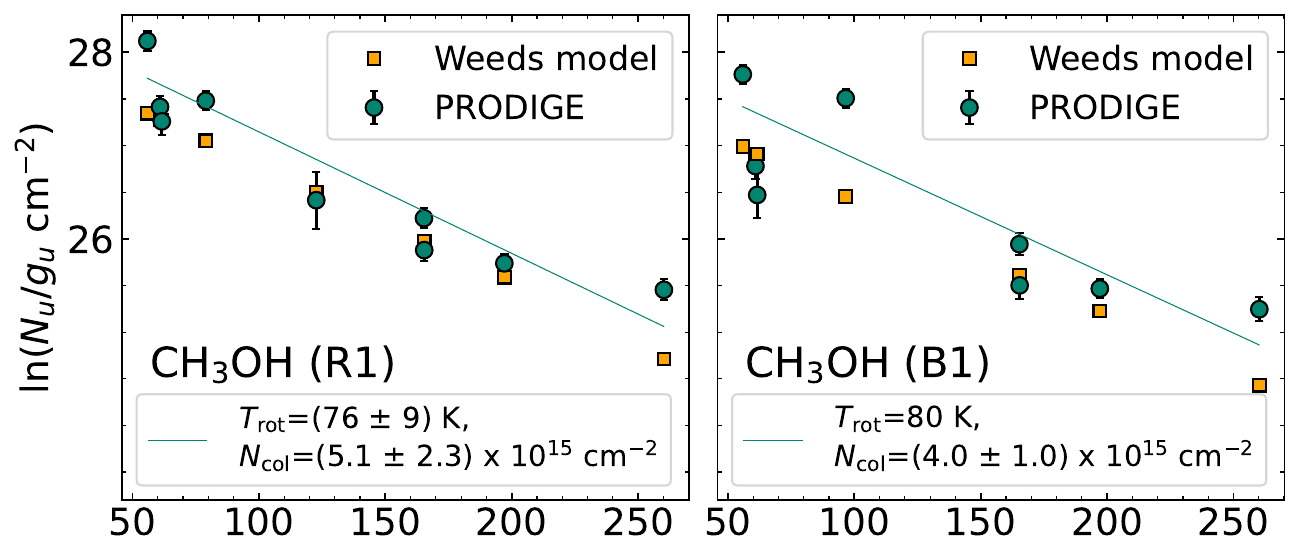}
    \includegraphics[width=0.47\textwidth]{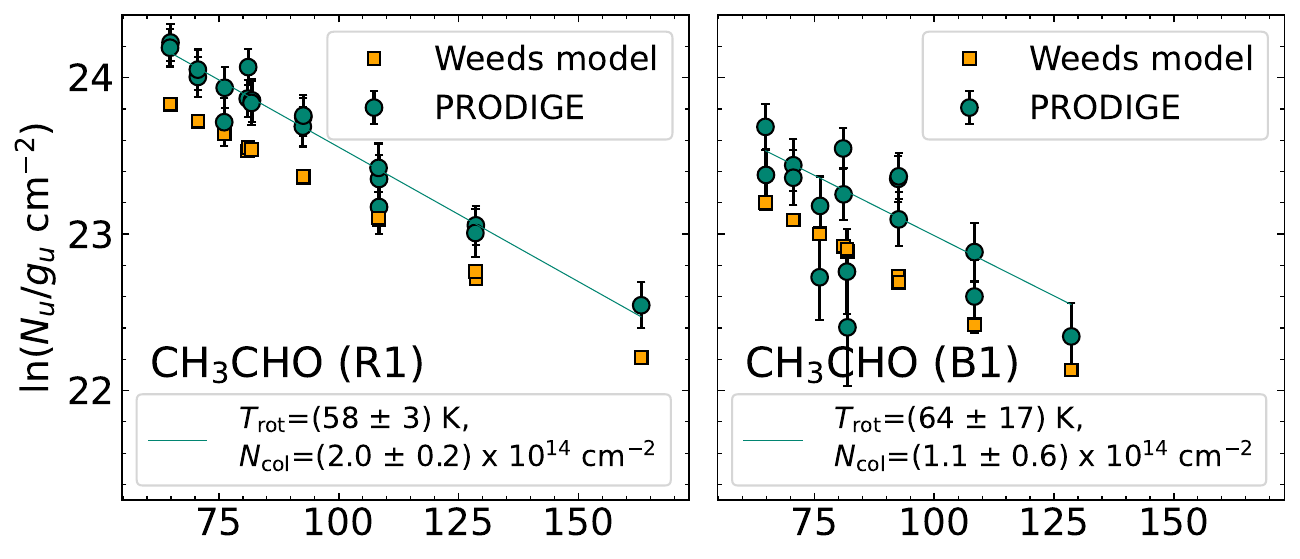}
    \includegraphics[width=0.47\textwidth]{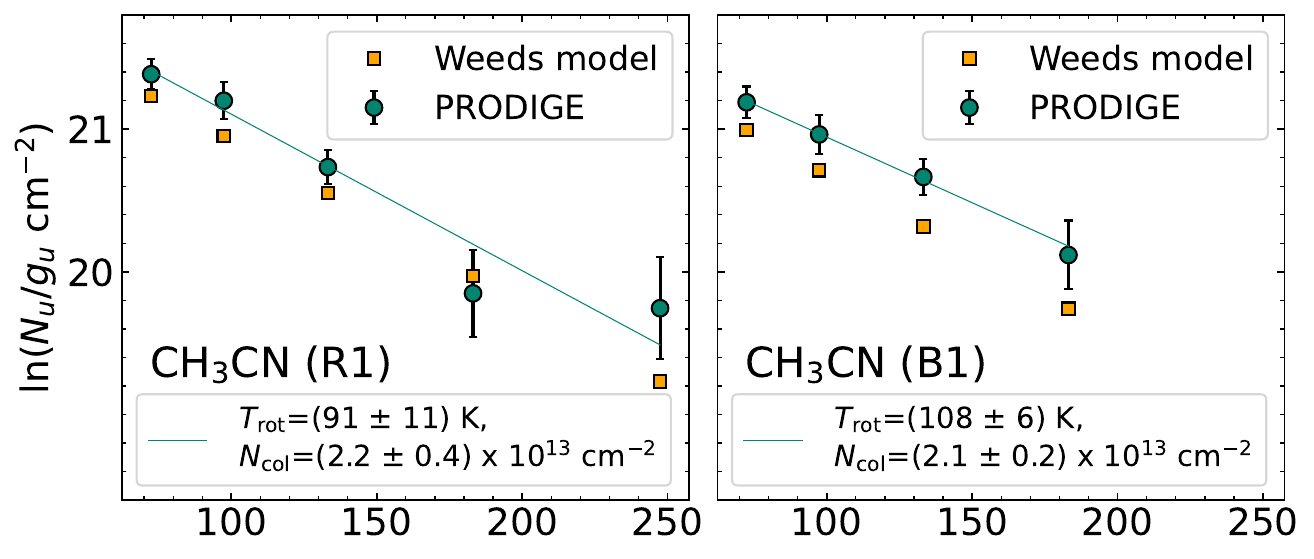}
    \includegraphics[width=0.47\textwidth]{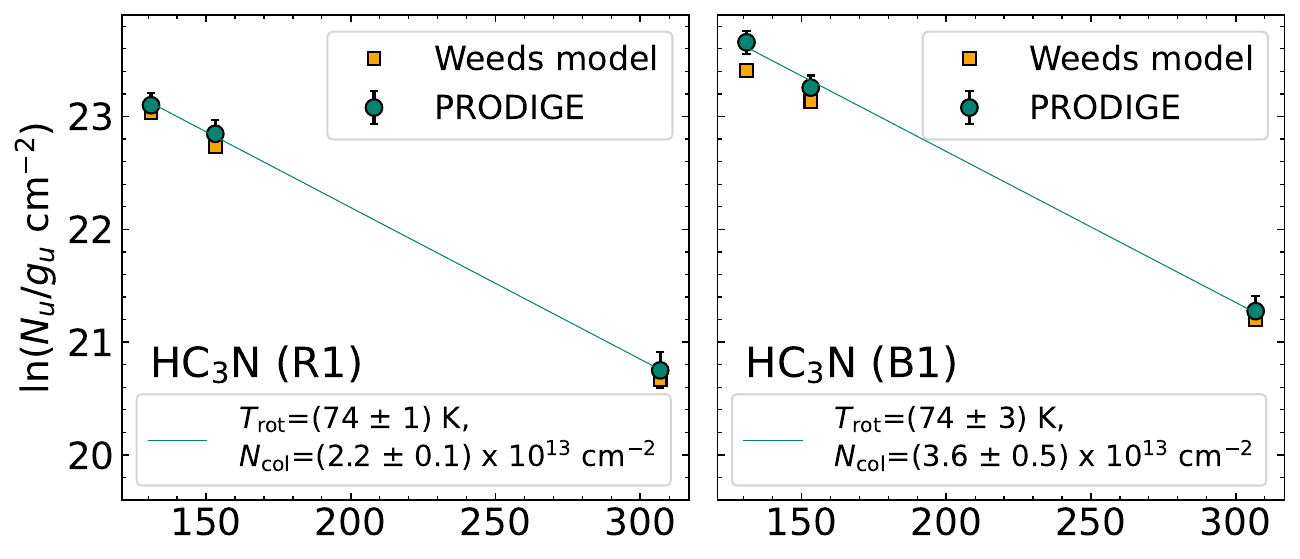}
    \includegraphics[width=0.47\textwidth]{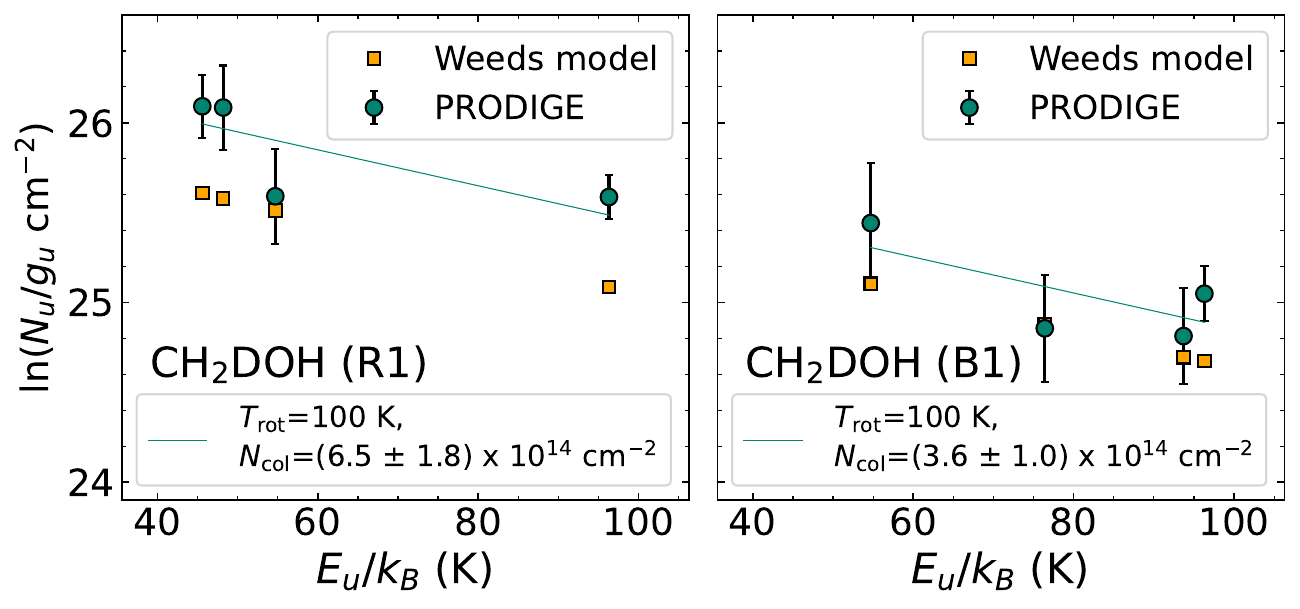}
    \includegraphics[width=0.47\textwidth]{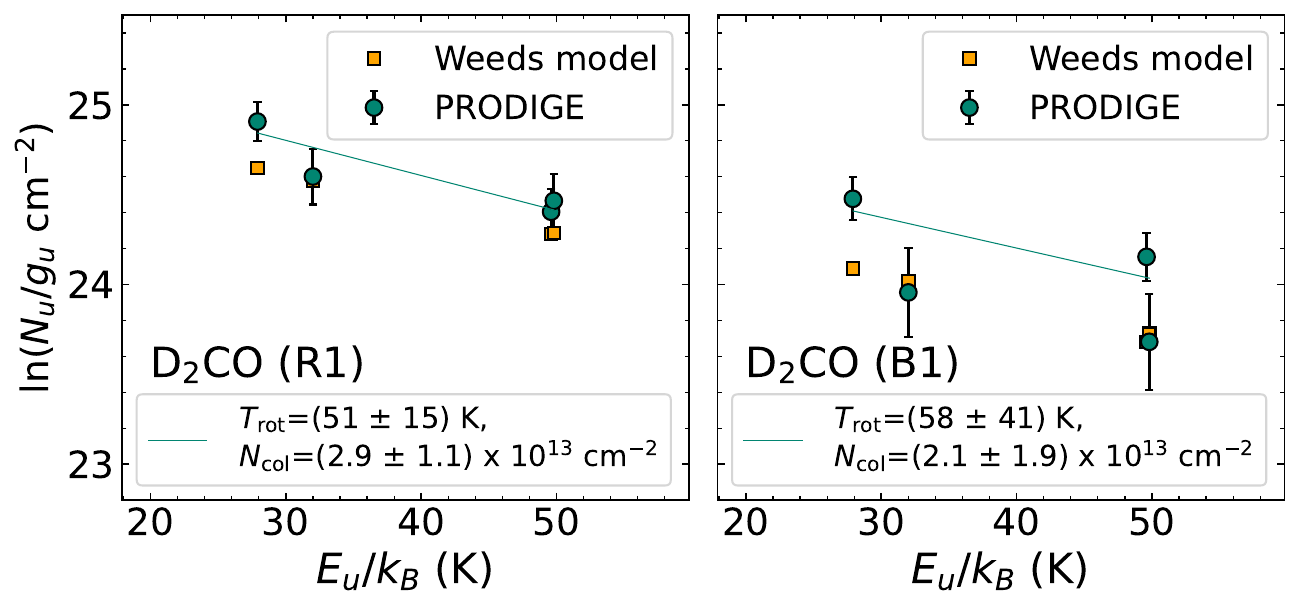}
    \caption{Population diagrams towards positions R1 and B1 (see Fig.\,\ref{fig:mom0}). The results of the linear fit to the observed data points (teal circles) are shown in the bottom. The observed data were corrected for contaminating emission, and both observed and modelled data (orange squares) are corrected for optical depth \citep[for details see][]{Busch2025}. }
    \label{fig:PD}
\end{figure}

Figure\,\ref{fig:spectra} shows example spectra towards positions R1 and B1 (Fig.\,\ref{fig:main}) for selected molecules. Vertical lines indicate the integration limits used for Figs.\,\ref{fig:main}, \ref{fig:ratios}, and \ref{fig:mom0}.
Figure\,\ref{fig:chan_maps} shows channel maps towards the IRAS\,4B Class 0 binary of SiO and CO using PRODIGE data and \met and C$^{34}$S using ALPPS data. Intensities were integrated from $-33$ to 47\,\kms in 10\,\kms intervals for SiO and CO and from 3 to 11\,\kms in 2\,\kms intervals for \met and C$^{34}$S. The ALPPS data at 3--4$\times$ higher angular resolution reveal the highly structured outflow morphology.
Figure \ref{fig:mom0} shows integrated intensity maps of red- and blueshifted emission from various organic molecules using the PRODIGE data. 
The derived population diagrams are shown in Fig.\,\ref{fig:PD}.

Table\,\ref{tab:lines} summarises spectroscopic properties of molecular transitions used for the various maps in this work.
Table\,\ref{tab:analysis} shows the results of the LTE spectral-line analyses towards positions IRAS\,4B1-B1 and IRAS\,4B1-R1 in the  outflow, using the radiative transfer code Weeds \citep[][]{Maret2011} and population diagrams \citep[][]{Goldsmith1999}. Column densities derived towards the L1157-B1 outflow shock spot are provided as well. Table\,\ref{tab:specprops} provides references for the spectroscopic information for the analysed molecules. 

\vspace{-.55cm}
\begin{figure*}[h]
    \centering
    \includegraphics[width=.93\linewidth]{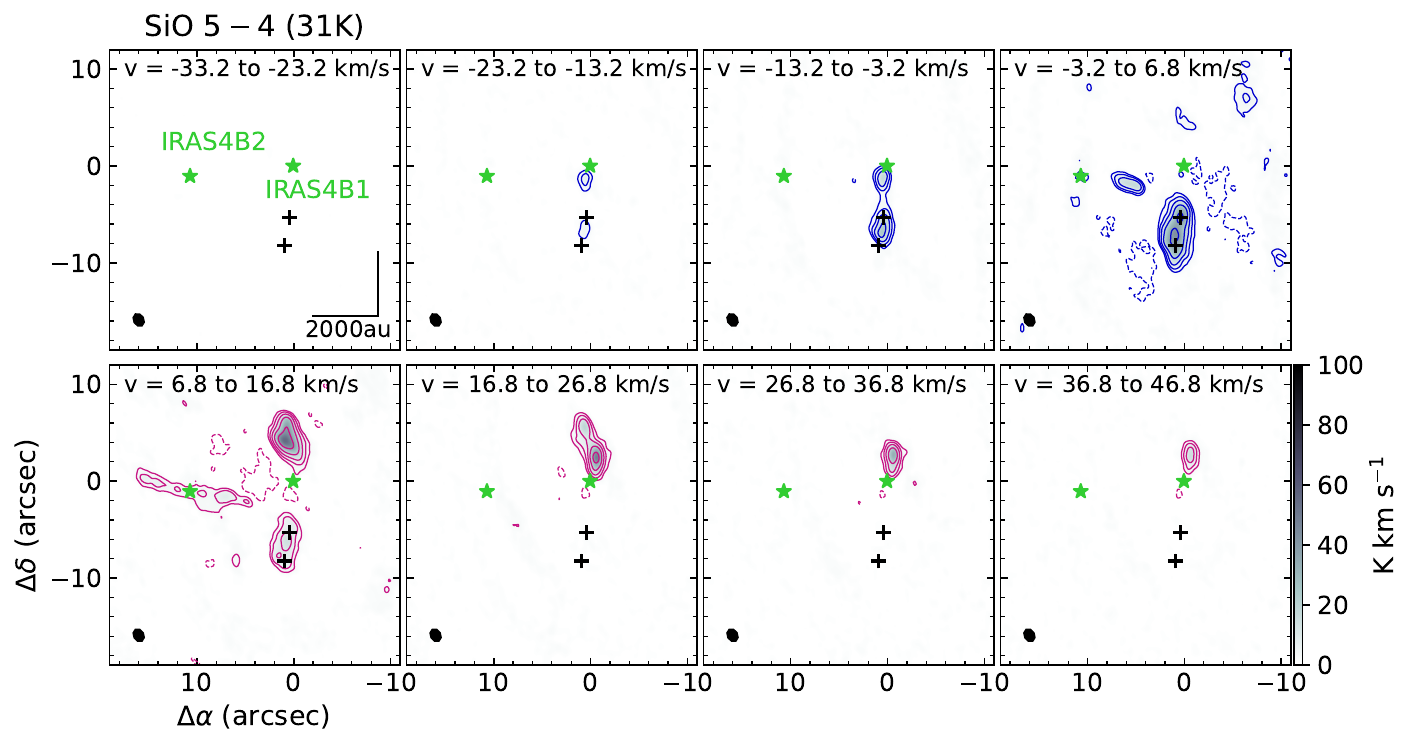}
    \vspace{-0.25cm}
    \caption{Integrated intensity maps for SiO and CO (PRODIGE) towards IRAS\,4B1 and IRAS\,4B2 (green stars) and for \met and C$^{34}$S (ALPPS, next page) zoomed in (cf. black dashed rectangle in CO panels) on IRAS\,4B1. Intensities were integrated over intervals of 10\,\kms (SiO and CO) and 2\,\kms (\met and C$^{34}$S). Contour levels are at $-10\sigma$, 10$\sigma$ and then increase by a factor 2, where $\sigma=0.24$\,K\,\kms (SiO), 0.27\,K\,\kms (CO), and 0.34\,K\,\kms (\met and C$^{34}$S). Black crosses mark H\2 knots \citep{Choi2011}. The HPBWs are shown in the bottom left.}
    \label{fig:chan_maps}
\end{figure*}
\begin{figure*}
    \addtocounter{figure}{-1}
    \centering
    
    \includegraphics[width=\linewidth]{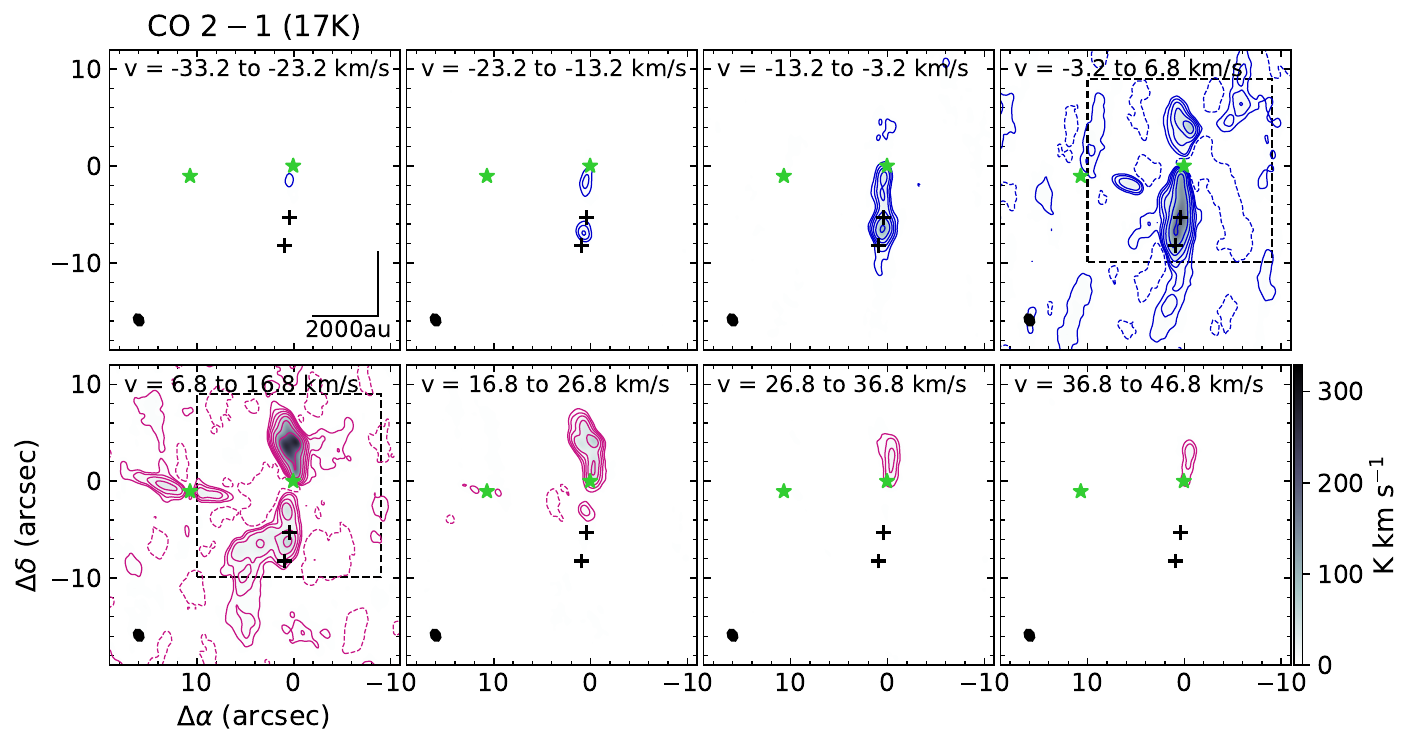}
   \includegraphics[width=.94\linewidth]{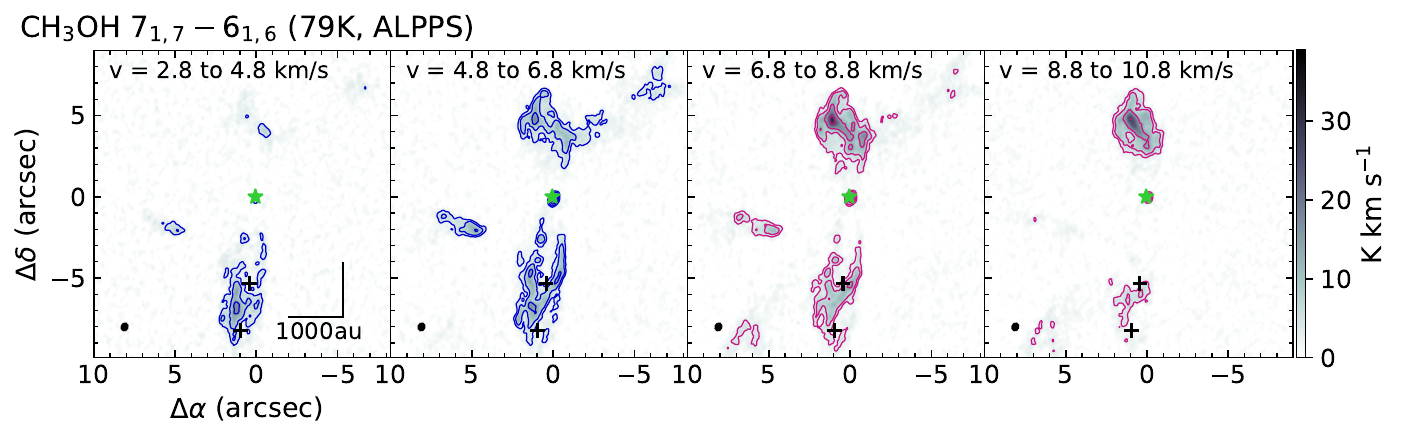}\vspace{-0.1cm}
    \includegraphics[width=.94\linewidth]{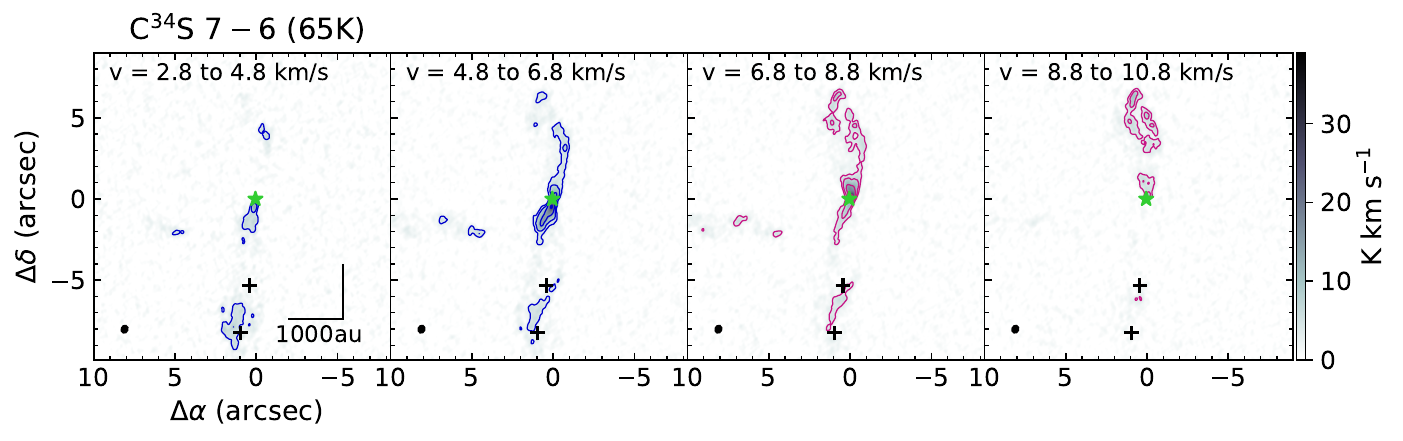}
    \caption{continued.}
\end{figure*}

%
%
%
%
\begin{table*}[h]
\caption{Molecular transitions used in this work for integrated intensity and temperature maps.} 
\centering
\begin{tabular}{lcccccccl}
\hline\hline \\[-0.3cm] 
Molecule & Transition & Frequency & $E_u$\tablefootmark{a} & $A_u$\tablefootmark{b} & $g_u$\tablefootmark{c} & $n_\mathrm{crit}$\tablefootmark{d} & Survey\tablefootmark{e} & Refs.\tablefootmark{f} \\ 
 & & (MHz) & (K) & (s$^{-1}$) & & (cm$^{-3}$) & & \\\hline \\[-0.3cm] 
CO & 2 -- 1 & 230538.000 & 16.6 & 6.9($-$7) & 5 & 1(4) & P & \cite{Yang2010-co}, \\
SiO & 5 -- 4 & 217104.919 & 31.3 & 5.2($-$4) & 11 & 3(6) & P & \cite{balanca2018-sio} \\
C$^{34}$S & 7 -- 6 & 337396.459 & 64.8 & 7.6($-$4) & 15 & 9(6) & A & \cite{Denis-Alpizar2018-cs}\tablefootmark{*} \\
$c$-C$_3$H$_2$ & $3_{3,0}-2_{2,1}$ & 216278.756 & 19.5 & 2.6($-$4) & 21 & 5(7) & P & \cite{Chandra2000-c3h2} \\
H$_2$CO & $3_{0,3}-2_{0,2}$ & 218222.192 & 21.0 & 2.8($-$4) & 7 & 3(6) & P & \cite{Wiesenfeld2013-h2co} \\
DCN & 3 -- 2 & 217238.612 & 20.9 & 4.6($-$4) & 9 & 4(7) & P & \cite{Dumouchel2010-hcn}\tablefootmark{*} \\
D$_2$CO & $4_{1,4}-3_{1,3}$ & 221191.661 & 32.0 & 2.8($-$4) & 9 & 4(6) & P & \cite{Wiesenfeld2013-h2co}\tablefootmark{*} \\
CH$_3$OH & $5_{1,4}-4_{2,2}$ & 216945.521 & 55.9 & 1.2($-$5) & 44 & 7(6) & P & \cite{Rabli2010-methanol} \\
 & $4_{2,2}-3_{1,2}$ & 218440.063 & 45.5 & 4.7($-$5) & 36 & 8(7) & P \\
 & $10_{2,9}-9_{3,6}$ & 231281.110 & 165.3 & 1.8($-$5) & 84 & 3(7) & P \\
 & $7_{1,7}-6_{1,6}$ & 335582.017 & 79.0 & 1.6($-$4) & 60 & 2(6) & A \\
 & $12_{1,11}-12_{0,12}$ & 336865.149 & 197.1 & 4.1($-$4) & 100 & 9(7) & A \\
CH$_3$CHO & $11_{1,10}-10_{1,9}$ & 216581.933 & 64.9 & 3.5($-$4) & 23 & -- & P \\
HC$_3$N & 24 -- 23 & 218324.723 & 131.0 & 8.3($-$4) & 49 & 2(7) & P & \cite{Faure2016-hc3n} \\
 & 37 -- 36 & 336520.084 & 306.9 & 3.1($-$3) & 75 & 1(7) & A \\
CH$_3$CN & $12_{3}-11_{3}$ & 220707.753 & 133.2 & 6.0($-$6) & 50 & 2(5) & P & \cite{Khalifa2023-ch3cn} \\
HNCO & $10_{1,10}-9_{1,9}$ & 219798.274 & 58.0 & 1.5($-$4) & 21 & 2(6) & P & \cite{Sahnoun2018} \\
\hline\hline
\end{tabular}
\tablefoot{Spectroscopic information are taken from the Leiden Atomic and Molecular Database \citep[LAMDA;][]{LAMDA2005}. \\ {$^{(a)}$}{Upper-level energy.} {$^{(b)}$}{Einstein A coefficient, where x(y) = x $\times$ 10$^{y}$.} {$^{(c)}$}{Statistical weight.} {$^{(d)}$}{Critical density estimate, $n_\mathrm{crit,T}=A_u/C_u$, for $T=100$\,K. The collisional rate coefficients are taken from the references in column nine.} {$^{(e)}$}{Survey designation: P = PRODIGE, A = ALPPS.} {$^{(f)}$}{References for collisional rate coefficients.} {$^{(*)}$}{Collisional rate coefficients are for the transition of the main isotopologues.}} 
\label{tab:lines} 

\caption{Weeds parameters and results of the population-diagram analysis. References to spectroscopic information used for the analysis are given in Table\,\ref{tab:specprops}.} 
\centering
\small
\begin{tabular}{lccccccr|c}
\hline\hline \\[-0.3cm] 
Molecule\tablefootmark{a} & P & $T_{\rm rot,W}$\tablefootmark{b} & $T_{\rm rot,PD}$\tablefootmark{c} & $N_{\rm tot,W}\tablefootmark{d}$ & $N_{\rm tot,PD}\tablefootmark{e}$ & $\Delta\varv$\tablefootmark{f} & $\varv_{\rm off}$\tablefootmark{g} & $N_{\rm L1157-B1}\tablefootmark{h}$  \\ 
 & & (K) & (K) & (cm$^{-2}$) & (cm$^{-2}$) & (km\,s$^{-1}$) & (km\,s$^{-1}$) & (cm$^{-2}$) \\\hline \\[-0.3cm] 
D$_2$CO & R1 & $60$ & $51\pm 15$ & $2.8(13)$ & $(2.9\pm 1.1)(13)$ & $4.5$ & $2.0$ &  \\[0.05cm] 
  & B1 & $60$ & $58\pm 41$ & $1.6(13)$ & $(2.1\pm 1.9)(13)$ & $4.5$ & $-0.5$ & \\[0.05cm] 
 CH$_3$OH & R1 & $80$ & $76\pm 9$ & $5.0(15)$ & $(5.1\pm 2.3)(15)$ & $5.0$ & $2.5$ & $(1.3\pm0.3)(16)$ \\[0.05cm] 
  & B1 & $80$ & $80$ & $3.5(15)$ & $(4.0\pm 1.0)(15)$ & $6.5$ & $-1.5$ & \\[0.05cm] 
 CH$_2$DOH & R1 & $100$ & $100$ & $4.5(14)$ & $(6.5\pm 1.8)(14)$ & $3.5$ & $1.5$ & $(2.2\pm0.7)(14)$ \\[0.05cm] 
  & B1 & $100$ & $100$ & $3.0(14)$ & $(3.6\pm 3.6)(14)$ & $4.5$ & $-1.5$ & \\[0.05cm] 
 CH$_3$CHO & R1 & $60$ & $58\pm 3$ & $1.5(14)$ & $(2.0\pm 0.2)(14)$ & $4.5$ & $2.5$ & $(7.0\pm3.0)(13)$ \\[0.05cm] 
  & B1 & $60$ & $64\pm 17$ & $8.0(13)$ & $(1.1\pm 0.6)(14)$ & $4.5$ & $-1.5$ & \\[0.05cm] 
 HC$_3$N & R1 & $75$ & $74\pm 1$ & $2.0(13)$ & $(2.2\pm 0.1)(13)$ & $6.0$ & $3.0$ & $3.4(13)$ \\[0.05cm] 
  & B1 & $80$ & $74\pm 3$ & $2.8(13)$ & $(3.6\pm 0.5)(13)$ & $7.0$ & $-2.5$ & \\[0.05cm] 
 CH$_3$CN & R1 & $90$ & $91\pm 11$ & $1.9(13)$ & $(2.2\pm 0.4)(13)$ & $5.5$ & $2.5$ & $1.0(13)$ \\[0.05cm] 
  & B1 & $90$ & $108\pm 6$ & $1.5(13)$ & $(2.1\pm 0.2)(13)$ & $5.5$ & $-1.5$ & \\[0.05cm] 
 NH$_2$CHO & R1 & $50$ &  & $ <1.5(13)$ &  & $5.0$ & $2.5$ & $8.0(12)$ \\[0.05cm] 
  & B1 & $50$ &  & $ <1.5(13)$ &  & $5.0$ & $-2.5$ & \\[0.05cm] 
 \hline\hline
\end{tabular}
\tablefoot{Values in parentheses show the decimal power, where $x(z) = x\times 10^z$ or $(x\pm y)(z) = (x\pm y)\times 10^z$. \\\tablefoottext{a}{Detected molecule and vibrational state used to derive population diagrams.}\tablefoottext{b}{Rotational temperature used for the Weeds model.} \tablefoottext{c}{Rotational temperature derived from the population diagram. When a value has no error, it was fixed.}\tablefoottext{d}{Column density used for the Weeds model. An upper limit on $N_\mathrm{tot,W}$ indicates that the molecule is not detected.} \tablefoottext{e}{Column density derived from the population diagram.}\tablefoottext{f}{$FWHM$ of the spectral lines.}\tablefoottext{g}{Offset from $\varv_\mathrm{sys}=6.8$\,\kms.}\tablefoottext{h}{Column densities derived from interferometric observations towards L1157-B1: \met and \ad \citep{Codella2020}}, HC\3N \citep{Benedettini2007}, \metc \citep{Codella2009}, CH\2DOH \citep{Fontani2014}, and \fmm \citep{Codella2017}.}
\label{tab:analysis} 
\end{table*}

\begin{table*}[h]
    \caption{Spectroscopic references for the analysed molecules in Table\,\ref{tab:analysis}.}
    \centering
    \begin{tabular}{lll}
    \hline\hline
        Molecule & DB\tablefootmark{a} & References \\\hline\\[-0.3cm]
        D\2CO & CDMS & \cite{Tucker1973-d2co}, \cite{Chardon1974-d2co}, \cite{Dangoisse1978-d2co}, \cite{Basakov1988-d2co}, \\
        & & \cite{Bocquet1999-d2co} \\
        \met & CDMS & \cite{Xu2008-met} \\
        CH\2DOH & JPL & \cite{Pearson2012-dmet} \\
        \ad & LSD & \cite{Smirnov2014-AD} \\
        HC\3N & CDMS & \cite{deZafra1971-hc3n}, \cite{Mallinson1978-hc3n}, \cite{DeLeon1985-hc3n}, \cite{Chen1991-hc3n} \\
        & & \cite{Yamada1995-hc3n}, \cite{Thorwirth2000-hc3n} \\
        \metc & CDMS & \cite{Mueller2015-ch3cn}, \cite{gadhi1995-ch3cn} \\
        \fmm & CDMS & \cite{Kurland1957-fmm}, \cite{Kukolich1971-fmm}, \cite{Hirota1974-fmm}, \cite{Gardner1980-fmm}, \\
        & & \cite{Moskienko1991-fmm}, 
        \cite{vorobeva1994-fmm}, \cite{Blanco2006-fmm}, \cite{Kryvda2009-fmm} \\
        \hline\hline
    \end{tabular}
    \label{tab:specprops}
    \tablefoot{{$^{(a)}$}{Database (DB): Cologne database for molecular spectroscopy \citep[CDMS;][]{CDMS}, Lille spectroscopic database \citep[LSD;][]{Motiyenko2025}, Jet Propulsion Laboratory (JPL) catalogue \citep[][]{JPL}}.}
\end{table*}

%

\end{document}